\documentclass[aps,onecolumn,10pt]{revtex4}
\usepackage{epsfig}
\usepackage{graphicx}
\usepackage{amsmath}
\usepackage{amssymb}
\usepackage{hyperref}
\hypersetup{colorlinks=true,linkcolor=red,citecolor=green,urlcolor=blue}
\usepackage{float}
\numberwithin{equation}{section}

\begin{document}
\allowdisplaybreaks
\setcounter{equation}{0}

\title{Critique of Breit-Wigner resonance scattering}

\author{Philip D. Mannheim}
\affiliation{Department of Physics, University of Connecticut, Storrs, CT 06269, USA \\
philip.mannheim@uconn.edu\\ }

\date{November 11 2026}

\begin{abstract}

In the Breit-Wigner approach to scattering the phase shift $\delta$ is to have a form $\tan\delta_{\rm BW} =\Gamma_1/(E_1-E)$ at any real energy continuum scattering resonance. By analytic continuation this leads to complex energy poles in the scattering amplitude at $E_{\rm BW}=E_1-i\Gamma_1$, poles that are identified with unstable physical particles. By solving the square well scattering problem we identify some challenges to this approach. We find that setting $\tan\delta_{\rm BW} =\Gamma_1/(E_1-E)$ is not always a good description of the real energy continuum scattering amplitude, that $\Gamma_1$ can be negative, that $E_{\rm BW}$ is not in  fact an energy eigenvalue, and thus cannot be identified with a physical particle. To address these issues we solve the square well Schr\"odinger problem and find explicit complex energy eigenvalues of the form $E_{PT}=E_2-i \Gamma_2$ due to the antilinear $PT$ (Parity, Time Reversal) symmetry  of the Hamiltonian. However, $E_2\neq E_1$ and $\Gamma_2\neq \Gamma_1$. As Hamiltonian   eigenstates with energy $E_{PT}=E_2-i \Gamma_2$ these states can be identified with physical particles and correspond to phase shifts of the form $\tan\delta_{PT} =\Gamma_2/(E_2-E)$. Then by analytic continuation to real energy they give the same real energy resonance shape as a standard Breit-Wigner. However, the determined positions and widths of these resonances should be associated with $E_2$ and $\Gamma_2$ rather than with $E_1$ and $\Gamma_1$. We thus recognize two forms of resonance in the real energy scattering amplitude $f(E)$, those associated with the continuum (the complex $E$ plane branch cut) and those due to poles in the complex $E$ plane. Only the latter can be associated with real particles. The continuum resonances are not associated with Breit-Wigner modes, and the number of such resonances correlates with the number of square well bound states. 

\end{abstract}
\maketitle

\section{Introduction}
\label{S1}

In scattering experiments there is a real energy continuum that is associated with a branch cut in the  complex energy plane scattering amplitude $f(E)$, with the  cut starting at a branch point at the real energy threshold. In  the standard Breit-Wigner approach to scattering experiments with real energy $E$ any resonance peaks that appear  in the cross section at real energy $E_1$ with width $\Gamma_1$ are associated with a scattering phase shift $\delta$ in the vicinity of the resonance of the Breit-Wigner  form $\tan\delta_{\rm BW} =\Gamma_1/(E_1-E)$, since this form leads to a phase shift equal to $\pi/2$ (or an odd multiple thereof) when $E=E_1$. An analytic complex energy plane continuation of this form for the phase shift in the scattering amplitude $f(E)=e^{i\delta(E)}\sin\delta(E)$  leads to poles in the scattering amplitude at $E_{\rm BW}=E_1-i\Gamma_1$, poles that are identified with physical particles.  This approach raises two immediate concerns. The first is whether a form such as $\tan\delta_{\rm BW} =\Gamma_1/(E_1-E)$ actually is a good description of the real energy scattering continuum $f(E)$ in the first place, and the second is whether the resulting pole in the scattering amplitude should in fact be identified with a physical particle or  just treated as an analytic continuation artifact. For such a pole to actually be a physical particle at $E_{\rm BW}=E_1-i\Gamma_1$ then $E_{\rm BW}$ should be an eigenstate of the scattering Hamiltonian. 

To address these concerns  we study a completely solvable model, namely a spherically symmetric, finite depth square well of width $a$ and depth $V_0$. In the model we can determine the energy dependence of the s-wave scattering continuum amplitude $f(E)$ in a closed form, viz. (see (\ref{3.3b}) below) 
\begin{align}
f(E)= \frac{ka\tan(Ka)-Ka\tan(ka)}{(ka\tan(Ka)+iKa)(\tan(ka)-i)},
\label{1.1x}
\end{align}
where $E=\hbar^2K^2/2m=\hbar^2k^2/2m+V_0$,  and find that in the resonance region it actually is not that well described by a Breit-Wigner form, especially in the real part of the scattering amplitude. Moreover, we find that the fitted $\Gamma_1$ need not even be positive, which would be required of a resonance that decays. Also we find that if we do attempt to describe the resonance region by a Breit-Wigner form, the resulting pole in the scattering amplitude is not in fact at an energy eigenvalue of the square well Hamiltonian.

However, we do nonetheless want to associate resonances with physical particles, and indeed this is how many unstable states have been discovered, especially in high energy scattering. To address this issue we solve the square well Schr\"odinger equation in the complex energy region,  and due to the $PT$  symmetry that the square well Hamiltonian possesses ($P$ is parity, $T$ is time reversal) we find exact eigenstates with energy eigenvalues  $E_{PT}=E_2-i\Gamma_2$, eigenvalues that are determined by the parameters in the square well Hamiltonian. For such eigenvalue modes we show that it explicitly is the case that $E_2 \neq E_1$ and $\Gamma_2\neq  \Gamma_1$, i.e., the $PT$ eigenvalue is not at the Breit-Wigner pole. If we set $\tan\delta_{PT}=\Gamma_2/(E_2-E)$ then  the scattering amplitude takes the form $f_{PT}(E)=\Gamma_2/(E_2-i\Gamma_2-E)$, to explicitly exhibit a pole at $E_{PT}=E_2-i\Gamma_2$. An analytic  continuation of $f_{PT}$ to real $E$ then exhibits a resonance in the cross section $\vert f(E)\vert^2$ at $E=E_2$ with width $\Gamma_2$. Such a resonance is not due to a peak in the  value of the scattering amplitude along the continuum branch cut (the Breit-Wigner case), but is instead due to a pole that lies below the branch cut.  Our study thus shows that it is invalid to fit cross-section resonances with a standard Breit-Wigner $\tan\delta_{\rm BW} =\Gamma_1/(E_1-E)$ phase shift as in (\ref{3.6b}), viz.
\begin{align}
f_{\rm BW}(E)=\frac{\Gamma_1}{E_1-i\Gamma_1-E},\qquad 
 \vert f_{\rm BW}(E)\vert^2&=\frac{\Gamma_1^2}{(E_1-E)^2+\Gamma_1^2}.
\label{1.2x}
\end{align}
%. 
However, it is valid to fit cross-section resonances as in (\ref{3.15d}), viz.
\begin{align}
f_{PT}(E)=\frac{\Gamma_2}{E_2-i\Gamma_2-E}, \qquad \vert f_{PT}(E)\vert^2=\frac{\Gamma_2^2}{(E_2-E)^2+\Gamma_2^2},
\label{1.3x}
\end{align}
 with a $PT$ symmetry  $\tan\delta_{PT} =\Gamma_2/(E_2-E)$ phase shift  instead.

Despite this disparity between the origins of the $PT$ resonance and the Breit-Wigner resonance, given their equivalent forms ($\tan\delta_{\rm BW} =\Gamma_1/(E_1-E)$, $\tan\delta_{PT} =\Gamma_2/(E_2-E)$), for every square well $PT$ theory cross section  resonance there is an effective Breit-Wigner form that fits it with an energy $E_2$ and a width $\Gamma_2$. However, since the Breit-Wigner and $PT$ resonance formulae coincide, what is recorded as a Breit-Wigner fit is actually a $PT$ fit, with the inferred position and width of each resonance still being valid for the associated particle, though what are thought to be the Breit-Wigner  $E_1$ and $\Gamma_1$ are actually the $PT$ $E_2$ and $\Gamma_2$. Our study thus justifies the use of what was  a presumed  Breit-Wigner form for the scattering cross section (since there were resonances), albeit with  $\tan\delta_{\rm BW} =\Gamma_1/(E_1-E)$ being replaced by  $\tan\delta_{PT} =\Gamma_2/(E_2-E)$. Our study thus establishes a new and quite central role for $PT$ symmetry in scattering (as generalized to $CPT$ in the relativistic case \cite{Mannheim2018a}), while also providing support for our general contention that it is antllinear symmetry rather than Hermiticity that should be taken as the guideline for quantum theory \cite{Mannheim2018a}.

Our paper is organized as follows. In Sec. \ref{S2} we present the standard Breit-Wigner approach. In Sec. \ref{S3}  we explore the square well system, showing that the standard Breit-Wigner model of the real energy continuum is not that good a model for the continuum. However, even if the continuum resonances are not associated with a Breit-Wigner shape or associated with real particles, the square well continuum $f(E)$ still has peaks when the phase shift goes through $\pi/2$ or an odd multiple thereof.  We find peaks in which $\tan\delta $ is negative at energies  lower than the peak energy and positive at energies  higher. Such peaks could not be described by a Breit-Wigner form as that would require that $\Gamma_1$ be negative. We show that the number of peaks that we do find correlates with the number of bound states that the square well can support. When the strength of the square well potential is such that there is only one bound state there is only one continuum resonance, and so on. 

Also in Sec. \ref{S3}  we show that there are exact complex energy solutions to the square well Schr\"odinger equation. Then in Sec. \ref{S4} we discuss our results from the perspective of $PT$ symmetry and its associated pseudo-Hermiticity. For modes of the  form $e^{i\bar{k}\cdot\bar{x}-iEt}$ with $E=\hbar^2k^2/2m+V_0$, we see that when $E=E_R+iE_I$ is complex then so is $k=k_R+ik_I$. When the real parts of both $k$ and $E$ are positive (above threshold outgoing waves with associated s-wave wave function $e^{ik_Rr-iE_Rt}/r$), their imaginary parts then have the same sign as each other ($E_I=\hbar^2k_Rk_I/m$). Thus the associated s-wave $e^{-k_Ir+E_It}/r$ factor in $e^{i\bar{k}\cdot\bar{x}-iEt}$ leads when $E_I$ and thus $k_I$ are negative to modes called Gamow vectors, viz. modes  that decay exponentially in time while growing exponentially in space. And not just in the $PT$ case but also in the Breit-Wigner case, with any analytic continuation of the energy requiring a continuation of the momentum. 

Because of its $PT$ symmetry, by time reversal the square well Hamiltonian also has  exact incoming eigenstates of the form  $e^{-ik_Rr-k_Ir-iE_Rt-E_It}/r$  ($E_I$ and $k_I$ again negative), modes that contribute in the time-reversed scattering amplitude. These modes grow exponentially in time and  just like the Gamow vectors also grow exponentially in space, and can thus not provide any spatial  exponential damping in overlaps with Gamow vectors. Specifically, both the $e^{ik_Rr-k_Ir-iE_Rt+E_It}/r$ and $e^{-ik_Rr-k_Ir-iE_Rt-E_It}/r$ modes are eigenstates of the $PT$  invariant square well Hamiltonian,  the former being a pole in the $f(E)=e^{i\delta(E)}\sin\delta(E)$ scattering amplitude and the latter being a pole in the time-reversed scattering amplitude. While the two modes contribute to different scattering amplitudes (scattering amplitudes only involve outgoing modes, a non-$PT$-symmetric set up), they nonetheless both lie in the eigenspace of the Hamiltonian and thus could overlap. 

In Sec. \ref{S5} we discuss how we can accommodate Gamow vectors that grow exponentially in space by introducing an outgoing anti-Gamow vector of the form  $e^{ik_Rr+k_Ir-iE_Rt-E_I/t}/r$ ($E_I$ and $k_I$ again negative), a vector  that is not an eigenstate of the square well Hamiltonian, but whose overlap with the Gamow vector does not grow exponentially in time or space. This anti-Gamow vector  can appear as a square-integrable test  function in a rigged Hilbert space or as the decay product produced by the decay of the Gamow vector. In Sec. \ref{S6} we present our conclusions, and show that when one takes into consideration the decay products that are produced by  $E_2-i\Gamma_2$ decays, in (\ref{5.12b}) one obtains two poles but only one resonance in the cross section, viz
\begin{align}
&f_{PT}(E)=\frac{\Gamma_2}{E_2-i\Gamma_2-E}-\frac{\Gamma_2}{E_2+i\Gamma_2-E}=\frac{2i\Gamma_2^2}{(E_2-E)^2+\Gamma_2^2},
\qquad \vert f_{PT}(E)\vert^2=\frac{4\Gamma_2^4}{[(E_2-E)^2+\Gamma_2^2)]^2}.
\label{1.4x}
\end{align}
This yields a resonance structure that  can be also fitted as one-pole Breit-Wigner resonances to a good degree, but with inferred $\Gamma_2$ widths that differ from the $\Gamma_1$ widths that would be inferred by actually treating the same resonances as Breit-Wigner resonances. It is the $\Gamma_2$ widths that should then be associated with unstable particles that are produced in scattering experiments. If a long-lived unstable state decays outside the scattering region  that produced it (a one-stage process) we would use the one-pole (\ref{1.3x})  to fit the cross section, while if it is short lived and decays inside the scattering region (a two-stage process) we would use the two-pole (\ref{1.4x}) instead. Thus while we cannot use an $f_{\rm BW}(E)$ Breit-Wigner form given in (\ref{1.2x}) to fit the real energy continuum $f(E)$ that lies on the branch cut in the complex energy plane, nonetheless the structure of the $E_2-i\Gamma_2$ poles that lie below the cut  still yield  Breit-Wigner shaped resonances, approximately in the short-lived case and exactly in the long-lived case.

In an appendix we discuss constraints on the imaginary component of the momentum, showing analytically that it must be negative for outgoing modes, to thereby lead  to exponential spatial growth. We also show that for ingoing modes there is also exponential spatial growth. A negative imaginary component of the momentum for outgoing modes above threshold uniquely entails that $\Gamma_2$ be positive, i.e., decay in time. Thus unlike a Breit-Wigner form for continuum resonance modes, which is not forced to be fitted with positive $\Gamma_1$,  complex $E_2-i\Gamma_2$  $PT$ modes are forced to have positive $\Gamma_2$. This is  just as we would want for the resonances that  $E_2-i\Gamma_2$ modes then lead to in real energy cross-sections, resonances that even though they are not Breit-Wigner continuum resonance modes, nonetheless can be fitted as Breit-Wigner modes to  a good degree.

\section{The Standard Breit-Wigner Approach}
\label{S2}

In the conventional treatment of the  quantum mechanics of a non-relativistic, real short range potential well one has bound states in the well and continuum scattering states above it.  In the Breit-Wigner scattering approach each real energy resonance that occurs in continuum scattering above the well is associated near the resonance energy with an energy-dependent phase shift of the form $\tan \delta_{\rm BW}=\Gamma_1/(E_1-E)$ with real $E_1$ and $\Gamma_1$, with $\delta$ being equal to $\pi/2$ (or an odd multiple thereof) when the energy $E$ is at resonance $E=E_1$. With a phase shift of this form the scattering amplitude and propagator behave as 
\begin{align}
f_{\rm BW}(E)= e^{i\delta_{\rm BW}}\sin\delta_{\rm BW}= \frac{\tan\delta_{\rm BW}}{1-i\tan \delta_{\rm BW}}=\frac{\Gamma_1}{E_1-i\Gamma_1-E},\qquad
D_{\rm BW}(E)=\frac{1}{E-E_1+i\Gamma_1}
=\frac{E-E_1-i\Gamma_1}{(E-E_1)^2+\Gamma_1^2},
\label{2.1}
\end{align}
with a pole at $E=E_1-i\Gamma_1$. To determine the associated time dependence we Fourier transform and obtain
\begin{align}
D_{\rm BW}(t)=\frac{1}{2\pi}\int _{-\infty}^{\infty}dE e^{-iEt}D_{\rm BW}(E)=\frac{1}{2\pi }\int _{-\infty}^{\infty}dE \frac{e^{-iEt}}{E-E_1+i\Gamma_1}.
\label{2.2a}
\end{align}
With the pole being in the lower half of the complex $E$ plane, and with $e^{-iEt}$ vanishing on the lower-half plane circle at infinity when $t>0$ while vanishing on the upper-half plane circle at infinity when $t<0$, contour integration gives the causal, forward in time propagator 
\begin{align}
D_{\rm BW}(t)=\frac{1}{2\pi }\int _{-\infty}^{\infty}dE \frac{e^{-iEt}}{E-E_1+i\Gamma_1}=-i\theta(t)e^{-iE_1t-\Gamma_1 t}.
\label{2.3a}
\end{align}
Thus  $f_{\rm BW}(E)$ and $D_{\rm BW}(t)$ only possess a decaying mode, with a time delay $\Delta T(E)=\hbar d\delta_{\rm BW}/dE$ \cite{Wigner1955} off resonance and on resonance of the respective forms
\begin{align}
\Delta T(E)=\hbar\frac{d\delta_{\rm BW}}{dE}=\frac{\hbar\Gamma_1}{[(E-E_1)^2+\Gamma_1^2]},\qquad \Delta T(E_1)=\frac{\hbar}{\Gamma_1}.
\label{2.4a}
\end{align}
This time delay is due to the fact that the wave is held by the well for a time $\Delta T(E_1)$. To determine what state the Breit-Wigner pole might correspond to we continue our previous study of  an exactly solvable model, the finite depth square well \cite{Mannheim2025a}.

\section{The Square Well }
\label{S3}
\subsection{General features}

To see the role of complex energies and momenta we recall the treatment of $s$-wave scattering of a particle of mass $m$ by a three-dimensional, spherically-symmetric, finite-depth square well potential $V(r<a)=0$, $V(r>a)=V_0$, where $V_0$ is real, positive and  finite and $a$ is the radius of the potential.  Inside the well the Schr\"odinger equation is $i\hbar\partial_t\psi(r<a,t)=-(\hbar^2/2m)\nabla^2\psi(r<a,t)$, and outside the well we have $i\hbar\partial_t\psi(r>a,t)=(-(\hbar^2/2m)\nabla^2+V_0)\psi(r>a,t)$. 
With $E=\hbar^2 K^2/2m$, $E-V_0=\hbar^2k^2/2m$, the radial $\ell=0$ wave functions are given by $R_0(r<a)=A \sin(Kr)/r$, $R_0(r>a)=B\sin(kr+\delta)/r$ with an energy-dependent phase shift $\delta$. Then by continuity of both the wave function and its derivative at $r=a$,  the s-wave phase shift $\delta$ acquired by the $\ell=0$ component of the outgoing wave is found to obey
\begin{align}
A \sin(Ka)=B\sin(ka+\delta), \qquad A K\cos(Ka)=Bk\cos(ka+\delta),\qquad K\tan(ka+\delta)=k\tan(Ka).
\label{3.1}
\end{align}
With $\beta=k\tan(Ka)/K$ and $ka=\alpha$, from the third relation in (\ref{3.1}) we obtain
\begin{align}
\sin\delta =\frac{\beta\cos\alpha-\sin\alpha}{(1+\beta^2)^{1/2}},\qquad \cos\delta =\frac{\beta\sin\alpha+\cos\alpha}{(1+\beta^2)^{1/2}}, \qquad \tan\delta=\frac{\beta-\tan\alpha}{\beta\tan\alpha+1}.\qquad \beta=\frac{\tan\delta +\tan\alpha}{1-\tan\alpha \tan \delta}.
\label{3.2}
\end{align}
Bound states that  lie in the well have real $0<E<V_0$, with  outgoing continuum scattering waves having positive, real $E>V_0$. Real energy resonances occur when $\tan\delta=\infty$, i.e., when $\beta\tan\alpha+1=0$.

In terms of the phase shift $\delta$ one constructs an above threshold,  real $E$ continuum scattering amplitude of the form 
\begin{align}
f(E)= \frac{\tan\delta}{1-i\tan\delta}=\frac{(\beta-\tan\alpha)}{(\beta+i)(\tan\alpha-i)}=\frac{ka\tan(Ka)-Ka\tan\alpha}{(ka\tan(Ka)+iKa)(\tan\alpha-i)}.
\label{3.3b}
\end{align}
According to (\ref{3.2}) the phase shift goes through $\pi/2$ (or an odd multiple thereof) when
\begin{align}
\beta=
\frac{ka\tan(Ka)}{Ka}=-\frac{1}{\tan(ka)}.
\label{3.4b}
\end{align}
Denoting the resonance value of $E$ as  $E_1$, then in terms of an arbitrary constant $\lambda$, using (\ref{3.2})  we can set
\begin{align}
\beta \tan(ka)+1=\lambda(E_1-E), \qquad \beta-\tan\alpha=\lambda\Gamma_1
\label{3.5b}
\end{align}
near resonance provided $\beta\tan(ka)+1$ vanishes linearly in $E_1-E$.
Denoting   the value of $\beta$ at $E=E_1=\hbar^2k_1^2/2m+V_0$ by  $\beta_1=k_1a\tan(K_1a)/K_1a=-1/\tan(k_1a)$ (here $k_1$ is the value of the momentum  $k$ at $E=E_1$), assuming a Breit-Wigner form for the phase shift, and using (\ref{3.3b}) and (\ref{3.5b}), then just as in (\ref{2.1})  we obtain
\begin{align}
\tan\delta_{\rm BW}&=\frac{\Gamma_1}{E_1-E},\qquad \Gamma_1=\frac{\beta_1}{\lambda} +\frac{1}{\lambda\beta_1}=\frac{\Gamma(k_1)}{\lambda},\qquad f_{\rm BW}(E)=\frac{\Gamma_1}{E_1-i\Gamma_1-E}=\frac{\Gamma_1(E_1-E) +i\Gamma_1^2}{(E_1-E)^2+\Gamma_1^2},
\nonumber\\
 \vert f_{\rm BW}(E)\vert^2&=\frac{\Gamma_1^2}{(E_1-E)^2+\Gamma_1^2},\qquad \vert f_{\rm BW}(E_1)\vert^2=1,
\label{3.6b}
\end{align}
with (\ref{3.6b}) defining $\Gamma(k_1)$ as $\lambda \Gamma_1$.
By comparing the exact continuum $f(E)$ as given in (\ref{3.3b}) with $f_{\rm BW}(E)$ as given in (\ref{3.6b}) we now check to see how reliable the Breit-Wigner formula for $f_{\rm BW}(E)$ actually is both at real $E$ and at $E=E_1-i\Gamma_1$.

\subsection{Implications for Breit-Wigner phase shift formula for real energy scattering}

\begin{figure}[H]
\centering
\includegraphics[scale=0.3]{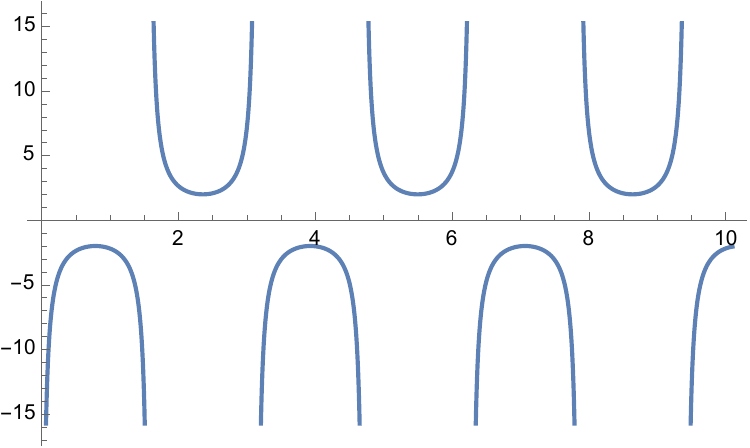}
\caption{$\Gamma(k)$ as a function of $k$.}
\label{gammak}
\end{figure}

In terms of any given arbitrary  momentum $k$ we first note that without any reference to the Breit-Wigner form we can determine $f(E)$ as given in (\ref{3.3b}) for any given $V_0$ as that would then fix $K$, with (\ref{3.2}) then fixing the phase shift $\delta$ as a function of energy. We thus want to compare $f(E)$ with $f_{\rm BW}(E)$ as a function of real $E$, i.e., real $k$. To determine the relevant $E_1$ and $\Gamma_1$ we fix $k_1$ and $K_1$ from (\ref{3.4b}) and $\Gamma_1$ from $\beta_1$. For a general $\Gamma(k)$ with general $k$ in Fig. \ref{gammak} we plot the function 
\begin{align}
\Gamma(k)=-\frac{1}{\tan(ka)}-\tan(ka)
\label{3.7c}
\end{align}
using the convenient simplification of setting both $\hbar^2/2m$ and $a$ to one, so that $K^2=E$, $k^2=E-V_0=K^2-V_0$. For every possible $k$ there is a value for $\Gamma(k)$, and as we see $\Gamma(k)$ is not actually required to be positive, and it could be negative. In terms of each such $k$  and any given $V_0$ we can then find an associated $K$, to then fix the energy dependence of the relevant $\delta$ via (\ref{3.2}).  

Since the overall sign of $\Gamma_1$ also depends on the sign of $\lambda$, to determine the sign  we have numerically studied the behavior of $\tan\delta$ as a function of $k$ in solutions to the square well theory that obey (\ref{3.2}) at some specific values of $V_0$, namely $V_0=4$, $V_0=8$ and $V_0=16$. 
\begin{figure}[H]
\centering
\includegraphics[scale=0.3]{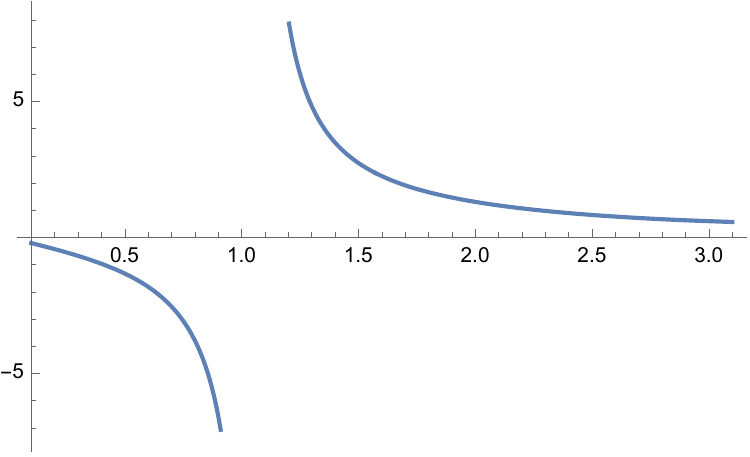}~~~\includegraphics[scale=0.3]{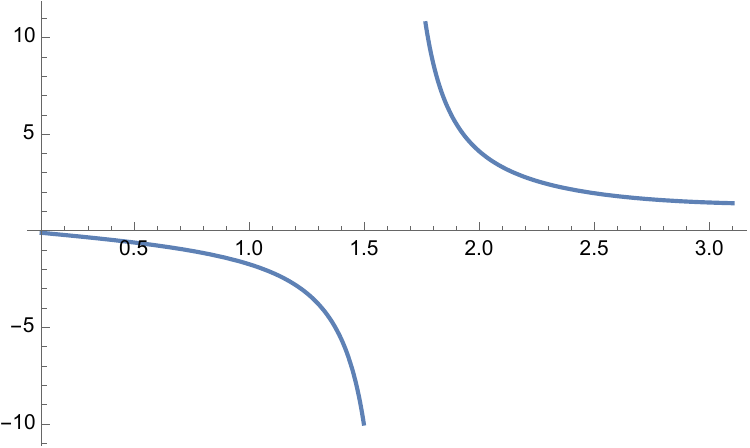}~~~
\includegraphics[scale=0.3]{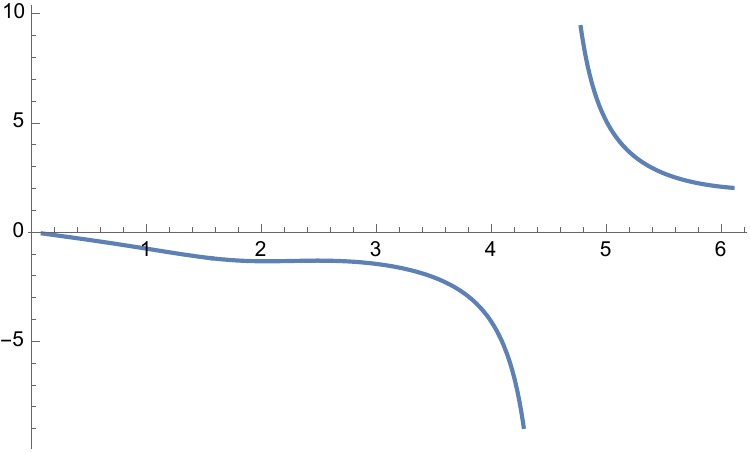}
\caption{$\tan\delta$  as a function of $k$ for $V_0=4$, $V_0=8$ and $V_0=16$.}
\label{tandelta}
\end{figure}

\smallskip

For $V_0=4$  we find that $\beta \tan k+1$ vanishes at $k=1.057$, at which point  $K=2.262$, $E=5.117$, $\beta_1=-0.564$, $\Gamma(1.057)=-2.336$. In  panel 1 in Fig.  \ref{tandelta} we plot $\tan\delta$ as a function of $k$ near to $k=1.057$, and see that it is negative in $k<1.057$ and positive in $k>1.057$. If parameterized as a Breit-Wigner, for $E<E_1$ and $k<1.057$ it follows that $\Gamma_1<0$. With $\Gamma(1.057)=-2.336$ it follows that the associated $\lambda$ is positive.

For $V_0=8$  we find that $\beta \tan k+1$ vanishes at $k=1.633$, at which point $K=3.266$, $E=10.667$, $\beta_1=0.062$, $\Gamma(1.633)=16.118$. In panel 2 in Fig.  \ref{tandelta} we plot $\tan\delta$ as a function of $k$ near to $k=1.663$, and see that it is negative in $k<1.633$ and positive in $k>1.633$.  If parameterized as a Breit-Wigner, for $E<E_1$ and $k<1.633$, it follows that $\Gamma_1<0$. Then with $\Gamma(1.633)=16.118$ it follows that the associated $\lambda$ is negative.

For $V_0=16$  we find that $\beta \tan k+1$ vanishes at $k=4.532$, at which point $K=6.045$, $E=36.539$, $\beta_1=--0.182$, $\Gamma(4.532)=-5.666$. In panel 3 in Fig.  \ref{tandelta} we plot $\tan\delta$ as a function of $k$ near to $k=4.532$, and see that it is negative in $k<4.532$ and positive in $k>4.532$.  If parameterized as a Breit-Wigner, for $E<E_1$ and $k<4.532$, it follows that $\Gamma_1<0$. Then with $\Gamma(4.532)=-5.666$ it follows that the associated $\lambda$ is positive.

\smallskip

All of the $V_0=4$, $V_0=8$, and $V_0=16$ $\tan\delta(k)$ plots show only one $\tan\delta(k)=\infty$ vertical asymptote, each one of which increases with $k$ from negative to positive, i.e., each one would correspond to a negative $\Gamma_1$. (There is nothing wrong with a $\tan\delta$ that goes from negative to positive across a vertical asymptote as that is the way the solution to the square well behaves. It is only  identifying it with a Breit-Wigner that would be wrong.) When continued to very large $k$ none of these $V_0=4$, $V_0=8$, and $V_0=16$ $\tan\delta(k)$ plots show any further $\tan\delta(k)=\infty$ asymptotes, with $\tan\delta(k)$ going to zero as $k\rightarrow \infty$. We have also found numerically that the smallest value of $V_0$ for which there is a vertical asymptote is $V_0=2.4674$. It occurs at $k= 0$, i.e., at threshold, and  if parameterized as a Breit-Wigner, it would have  $\Gamma_1$ negative. For there to be any square well bound states the potential has to satisfy $V_0>\hbar^2\pi^2/8ma^2$. With $a$ and  $\hbar^2/2m$ set equal to one this then requires that $V_0>\pi^2/4=2.4674$, so for this minimum value for $V_0$ the bound state and the vertical asymptote respectively lie just below and just above  threshold. Exactly at threshold in the $V_0=\hbar^2\pi^2/8ma^2$ case we have shown \cite{Mannheim2025a} that the threshold becomes an exceptional point at which the Hamiltonian becomes of non-diagonalizable Jordan-block form and has an incomplete set of eigenstates, with there being non-stationary modes that grow linearly in time. Thus in the particular case when $V_0=\hbar^2\pi^2/8ma^2$ there is a bound state when $E$ is slightly less than $V_0$, a continuum resonance when $E$ is slightly larger than $V_0$, and an exceptional point at the transition at $E=V_0$.

The condition that there be a second square well bound state is that  $V_0>9\hbar^2\pi^2/8ma^2\sim 9\pi^2/4$. We have found that the condition that there be  two vertical asymptotes is that $V_0>22.2066$, and $9\pi^2/4$ is equal to $22.2066$. Thus conditions that there be continuum modes with $\tan\delta=\infty$ completely track conditions for bound states. As a check we  found that for $V_0=24$ there indeed are two vertical asymptotes. They occur at $k=0.979$ and $k=7.326$, and  if parameterized as a Breit-Wigner, both have $\Gamma_1$ negative. Thus all of the modes that we have found numerically  are such that as $ k$ increases  $\tan\delta(k)$ goes from negative to positive, regardless of whether or not we parametrize $\tan\delta(k)$ as a Breit-Wigner. However, if we do parametrize $\tan\delta(k)$ as a Breit-Wigner, then in each of these cases  $\Gamma_1$ would be negative. This is problematic for the Breit-Wigner approach since if $\Gamma_1$ is negative we would get exponential growth in time in (\ref{2.3a}), something not allowed for in the Breit-Wigner approach since it is only designed for  decays (i.e., in $\tan\delta_{\rm BW}=\Gamma_1/(E_1-E)$ the parameter $\Gamma_1$ is required to be positive). Moreover, Breit-Wigner cross section fits using (\ref{3.6b}), viz. $\vert f_{\rm BW}(E)\vert^2=\Gamma_1^2/[(E_1-E)^2+\Gamma_1^2]$ , would not be sensitive to  the sign of $\Gamma_1$ since the cross section only depends on $\Gamma_1^2$, and one could thus inadvertently take $\Gamma_1$ to be positive in cases where the scattering Hamiltonian requires it to be negative. However, as we now show, this problem is moot since the Breit-Wigner formalism is not a good approximation to the continuum scattering amplitude in the first place, even if there are continuum resonances. 

\begin{figure}[H]
\centering
\includegraphics[scale=0.23]{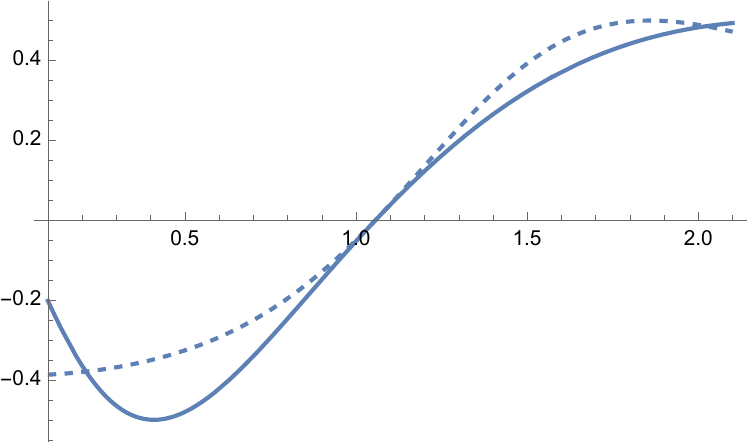}~~~\includegraphics[scale=0.23]{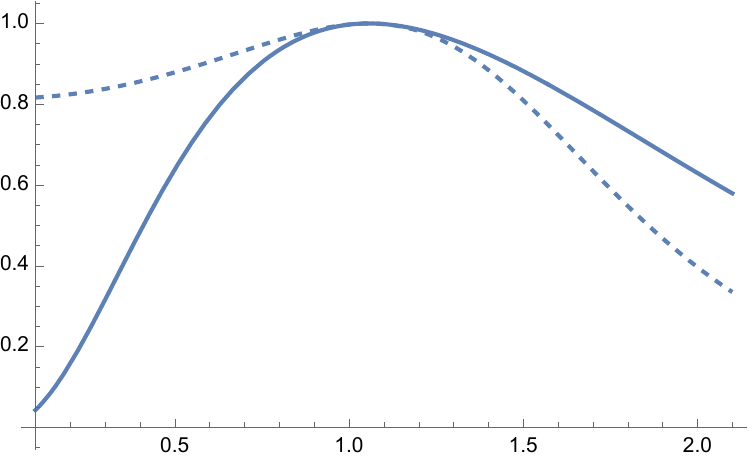}~~~
\includegraphics[scale=0.23]{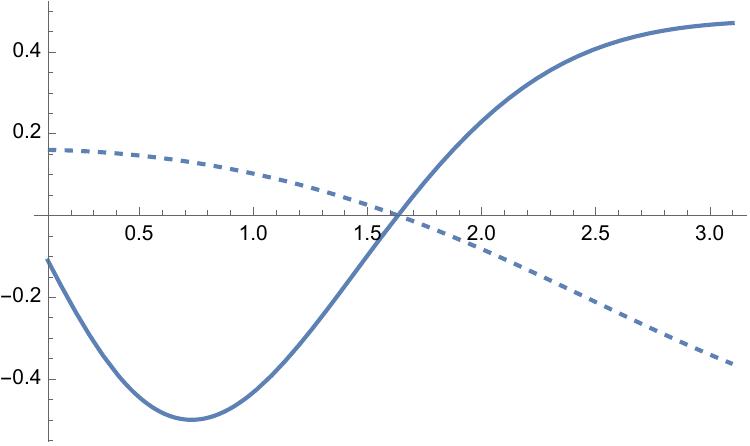}~~~\includegraphics[scale=0.23]{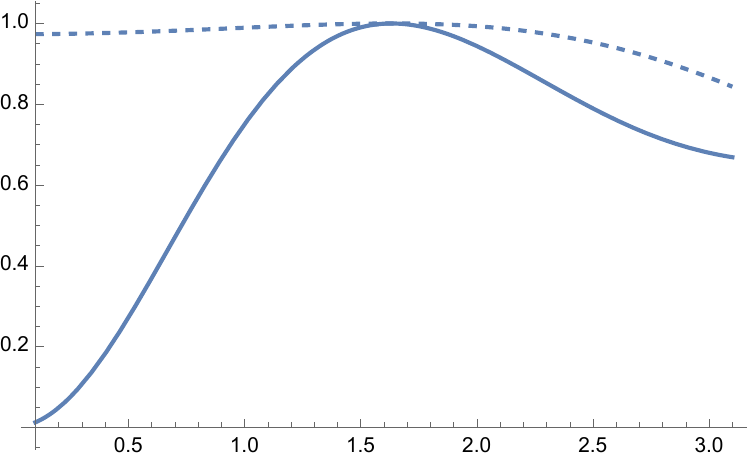}~~~
\includegraphics[scale=0.23]{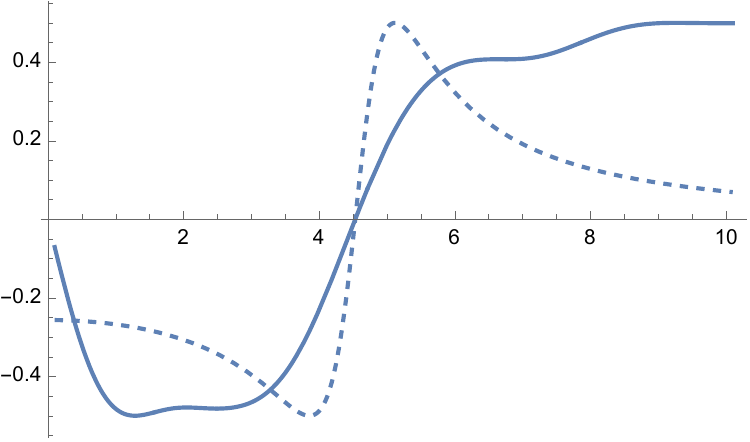}~~~\includegraphics[scale=0.23]{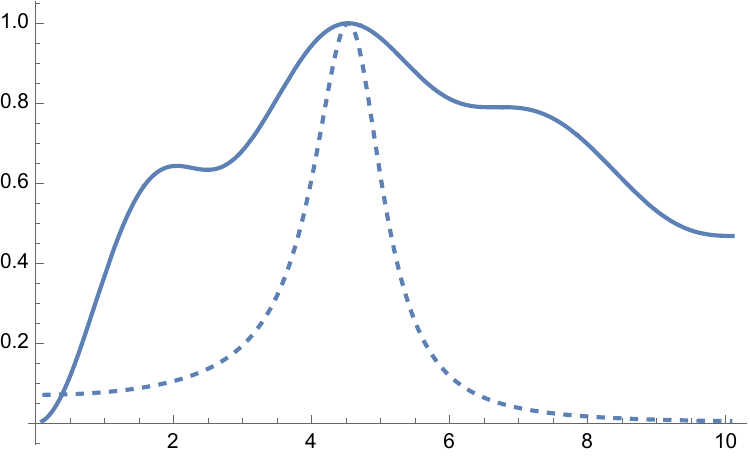}~~~
\caption{Exact real $f(E)$ (solid curve) and real $f_{\rm BW}(E)$ (dashed curve) in panels 1, 3, 5, exact imaginary $f(E)$ (solid curve) and imaginary $f_{\rm BW}(E)$ (dashed curve) in panels 2, 4, 6 as a function of $k$ plotted in sequence for $V_0=4$, $V_0=8$ and $V_0=16$.}
\label{ReIm}
\end{figure}
To see how good an approximation the Breit-Wigner form might be we compare the exact square well form for $f(E)$ given in (\ref{3.3b}) with the Breit-Wigner form given in (\ref{3.6b}), a form in which the parameter $\lambda$ drops out. In Fig. \ref{ReIm} we plot the real parts of $f(E)$ and $f_{\rm BW}(E)$ (panels 1, 3, 5) and then the imaginary  parts of $f(E)$ and $f_{\rm BW}(E)$ (panels 2, 4, 6) in sequence for our three $V_0=4$, $V_0=8$, and $V_0=16$ cases, with all real parts vanishing when each relevant  $\beta \tan(ka)+1$ is zero. Since the scattering amplitude is of the form $e^{i\delta}\sin\delta$ it follows that ${\rm Im}[f(E)]=\vert f(E)\vert ^2$, ${\rm Im}[f_{\rm BW}(E)]=\vert f_{\rm BW}(E)\vert ^2$ when $\delta$ is real, and so we have no need to plot $\vert f(E)\vert^2$ or $\vert f_{\rm BW}(E)\vert^2$ separately.  For the $V_0=8$ case (the one with negative $\lambda$) we see that there is a quite significant  difference for the real parts (even right on top of the resonance), while for the imaginary parts in the $V_0=4$ and $V_0=8$ cases the equivalence is only approximate, while in the $V_0=16$ case it is not good at all. There would have to be some difference for the imaginary part since $f(E)$ goes to zero at $k=0$ while $f_{\rm BW}(E)$ does not, though the main cause for concern is in the vicinity of the maximum where $\tan\delta$ is infinite, as agreement that is only valid for a very small spread around the maximum. Also we note that since $\vert f_{\rm BW}(E)\vert ^2$ does not depend on the sign of $\Gamma_1$, we cannot use the cross-section to determine the sign of $\Gamma_1$. However, we can use the real part of $f_{\rm BW}(E)$, and it is measurable in an interference experiment. This is of significance since the real part of the amplitude shows quite a significant departure of the Breit-Wigner form from  the exact form. 

\begin{figure}[H]
\centering
\includegraphics[scale=0.3]{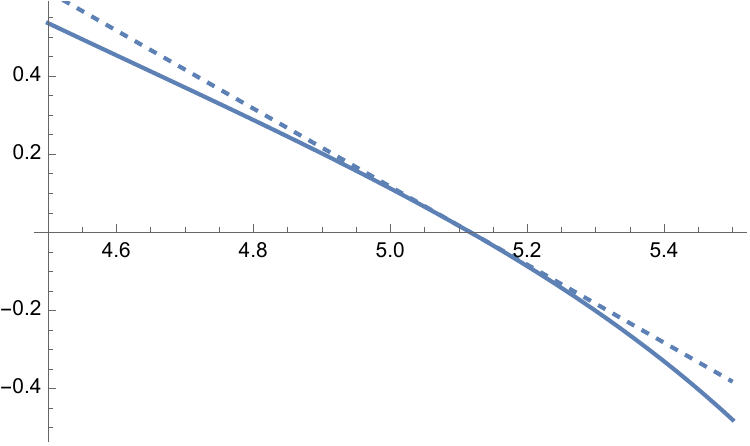}~~~\includegraphics[scale=0.3]{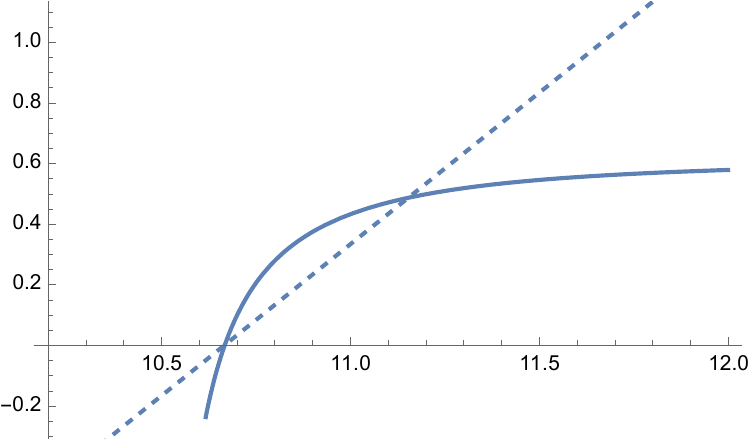}~~~
\includegraphics[scale=0.3]{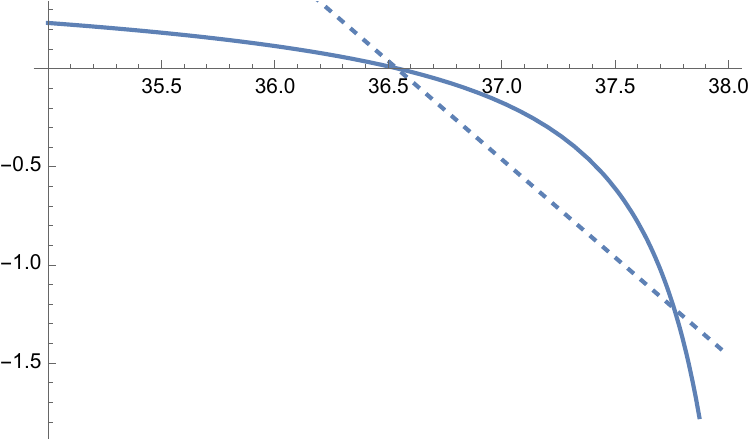}
\caption{Exact $\beta \tan(ka)+1$ (solid curve) and linear approximation $E_1-E$ (dashed curve) as a function of $E$ for $V_0=4$, $V_0=8$ and $V_0=16$.}
\label{linear}
\end{figure}

We can also check (\ref{3.5b}) to see whether the departure of $\beta\tan(ka)+1$ from  the point where it vanishes is in  fact linear in $E_1-E$. (If it were not then the only way we could have $\tan\delta_{\rm BW}=\Gamma_1/(E_1-E)$ would be if the $\Gamma _1$ numerator had a compensating dependence on $E_1-E$.) In Fig. \ref{linear}  we plot $\beta\tan(ka)+1$ and ${\rm sign}[\lambda](E_1-E)$ in the $V_0=4$, $V_0=8$, and  $V_0=16$ cases,  with $\lambda>0$ in the $V_0=4$ and $V_0=16$ cases, and $\lambda<0$ in the $V_0=8$ case.  As we see, for $V_0=8$ and $V_0=16$ the quantity $\beta\tan(ka)+1$ is not that well approximated as being linear in $E_1-E$ near the resonance, though we do  see that in each case $\lambda$ is of order one.

\subsection{Breit-Wigner in the complex energy plane}

When continued into the complex energy plane the Breit-Wigner form would give poles in the scattering amplitude at $E_1-i\Gamma_1$. We now show that these poles are not eigenvalues of the square well Schr\"odinger equation.

To study the structure of eigenstate solutions to (\ref{3.1}) we note that
\begin{align}
\frac{\sin(kr+\delta)}{r}=\frac{\cos\delta}{2r}[e^{ikr}(\tan\delta -i)+e^{-ikr}(\tan\delta+i)].
\label{3.8c}
\end{align}
Solutions to (\ref{3.8c}) that are to be outgoing $h_0^{(1)}(kr)=-ie^{ikr}/kr$ Hankel function modes (viz. real part of $k$ positive)  have to obey $\tan\delta+i=0$.
Solutions to (\ref{3.8c}) that are to be poles in the scattering amplitude $f(E)=\tan\delta/(1-i\tan\delta)$ also  have to obey $\tan\delta+i=0$, so that for both outgoing modes and poles  we have
\begin{align}
\frac{\sin(kr+\delta)}{r}=-\frac{i\cos\delta e^{ikr}}{r}.
\label{3.9c}
\end{align}
(Ingoing $e^{-ikr}/r$ modes ($h_0^{(2)}(kr)=ie^{-ikr}/kr$ Hankel functions) are associated with $\tan\delta -i=0$, and are poles in the time reversed scattering amplitude.) From (\ref{3.3b}) it  follows that the outgoing $\tan\delta+i=0$  poles obey $\beta=-i$. Thus we have
\begin{align}
\beta=
\frac{ka\tan(Ka)}{Ka}=-i,\qquad \cos(Ka)=\frac{k}{(k^2-K^2)^{1/2}},\quad \sin(Ka)=-\frac{iK}{(k^2-K^2)^{1/2}},
\label{3.10c}
\end{align}
with energy eigenvalues thus obeying 
\begin{align}
&\sin(Ka)=\left(\frac{E}{V_0}\right)^{1/2},\qquad \tan(Ka)=-\frac{iK}k=-i\left(\frac{E}{E-V_0}\right)^{1/2},
\nonumber\\
&\sin(\gamma(\mu+i\nu))=\frac{(\mu+i\nu)}{V_0^{1/2}},\qquad \tan(\gamma(\mu+i\nu))=-\frac{i(\mu+i\nu)}{[(\mu+i\nu)^2-V_0]^{1/2}},
\label{3.11c}
\end{align}
where for convenience we have set $E=(\mu+i\nu)^2$,  $(2m/\hbar^2)^{1/2}a=\gamma$, $Ka=\gamma(\mu+i\nu)$.
From the sine equation we can immediately find the standard bound state solutions  (viz.  states with $k=i\sigma$ where $\sigma$ is real and positive), since $\sin(Ka)$ and $E/V_0$ can both be less than one simultaneously. Above threshold there can only be solutions to (\ref{3.11c}) if $E$ and $K$ are complex, and note first that while the sine equation admits of solutions for either sign for $\nu$, the tangent equation only allows for one sign of $\nu$. To show that there actually are non-trivial complex solutions above threshold, from the sine equation  we obtain
\begin{align}
&\sin(\gamma\mu)\cosh(\gamma\nu)=\frac{\mu}{V_0^{1/2}},\quad \cos(\gamma\mu)\sinh(\gamma\nu)=\frac{\nu}{V_0^{1/2}},
\nonumber\\
&\cosh[\gamma[\mu^2\cot^2(\gamma\mu)-V_0\cos^2(\gamma\mu)]^{1/2}]=\frac{\mu}{V_0^{1/2}\sin(\gamma\mu)},\qquad 
\cos[\gamma[V_0\cosh^2(\gamma\nu)-\nu^2\coth^2(\gamma\nu)]^{1/2}]=\frac{\nu}{V_0^{1/2}\sinh(\gamma\nu)}.
\label{3.12c}
\end{align}
From (\ref{3.12c})  it follows that
\begin{align}
\frac{\tan(\gamma \mu)}{\gamma\mu}=\frac{\tanh(\gamma \nu)}{\gamma \nu},
\label{3.13c}
\end{align}
provided that  $\nu\neq 0$.
The relations given in (\ref{3.12c}) and (\ref{3.13c}) can then be solved for the two variables $\mu$ and $\nu$. Noting that these relations  are invariant under $\nu\rightarrow -\nu$, this then gives not one but two solutions for $E$ of the form 
\begin{align}
E_2\pm i\Gamma_2=\mu^2-\nu^2\pm 2i\mu\nu, \qquad  \mu^2=\frac{(E_2^2+\Gamma_2^2)^{1/2}+E_2}{2}, \qquad \nu^2=\frac{(E_2^2+\Gamma_2^2)^{1/2}-E_2}{2},
\label{3.14c}
\end{align}
solutions that are not just complex but in a complex conjugate pair. We note that $V_0$ is real and we have not given the potential any imaginary part.  

In regard to these solutions  we note that  for $\gamma\nu$ positive the quantity $\tanh(\gamma \nu)/\gamma \nu $ always lies between one and zero.  For $0<\gamma\mu<\pi/2$ the quantity  $\tan(\gamma \mu)/\gamma\mu $ is always  greater than one. For $\pi/2<\gamma\mu<\pi$ the quantity $\tan(\gamma \mu)/\gamma\mu $ is always negative. And since for $\pi<\gamma\mu<3\pi/2$ the quantity  $\tan(\gamma \mu)/\gamma\mu $ is not only positive but varies continuously between zero and infinity, there  could be a potential $\nu \neq 0$ solution to (\ref{3.13c}) (the first such one)  in the range $\pi < \gamma \mu <3\pi/2$, to thus be well above the threshold at $E=V_0$.  The next potential solution lies in the region $2\pi < \gamma \mu <5\pi/2$, and so on. (We refer to these solutions as being potential since we would still need to satisfy the other equations in (\ref{3.11c}) and (\ref{3.12c}) as well.) Since $\tanh(\gamma \nu)/\gamma \nu $  always lies between one and zero for either sign of $\nu$ any solutions to the sine equation would appear in pairs.

To show that there actually are  non-trivial solutions that we could then compare with the Breit-Wigner form we solve (\ref{3.12c})  and (\ref{3.13c}) for $V_0=4$, $V_0=8$ and  $V_0=16$, and set $\gamma=1$ for simplicity. In each case we report the lowest positive $\mu$ solution in each case.

\smallskip

For $V_0=4$ we find that $\mu=7.582$, $\nu=\pm 2.047$, $E_2=53.295$, $\Gamma_2=\mp 31.036$. 

For $V_0=8$ we find that $\mu=7.622$, $\nu=\pm1.677$, $E_2=55.285$, $\Gamma_2=\mp 25.569$. 

For $V_0=16$ we find that $\mu=7.661$, $\nu=\pm1.289$, $E_2=57.025$, $\Gamma_2=\mp19.742$. 

\smallskip

We note that all of these solutions lie in the allowed $2\pi \leq \mu\leq 5\pi/2$ range, and for all of them $E_2$ is above the $V_0$ threshold.  Since $K=\mu+i\nu$, we can identify $\mu$ and $\nu$ as the real and imaginary parts of $K=K_R+iK_I$, with $k=k_R+ik_I$ being given by  $k=[(\mu+i\nu)^2-V_0]^{1/2}$. 

Turning now to the tangent equation  we evaluate the quantity $T(\mu,\nu,V_0)=\tan(\gamma(\mu+i\nu))+i(\mu+i\nu)/[(\mu+i\nu)^2-V_0]^{1/2}$, which is required to be zero. 

\smallskip

For $V_0=4$ we find that $T(7.582,+2.047,4)=0.036+2.058 i$, $T(7.582,-2.047,4)=0$.

For $V_0=8$ we find that $T(7.622,+1.677,8)=0.067+2.128 i$, $T(7.622,-1.677,8)=0$.

for $V_0=16$ we find that $T(7.661,+1.289,16)=0.132+2.299 i$, $T(7.661,-1.289,16)=0$. 

\smallskip 

To determine the momentum we write (\ref{3.11c}) in the form $k=-iK/\tan(K)=-i(\mu+i\nu)\cot(\mu+i\nu)$, and evaluate $COT(\mu,\nu,V_0)=-i(\mu+i\nu)\cot(\mu+i\nu)$, which is required to be equal to $k$.

\smallskip

For $V_0=4$ we find that $COT(7.582,+2.047,4)=-7.333-2.116 i$, $COT(7.582,-2.047,4)=7.333-2.116 i$. 

for $V_0=8$ we find that $COT(7.622,+1.668,8)=-7.098-1.792 i$, $COT(7.622,-1.668,8)=7.098-1.792i$.

for $V_0=16$ we find that $COT(7.661,+1.289,16)=-6.580-1.500i$, $COT(7.661,-1.289,16)=6.580-1.500i$. 

\smallskip 

Thus in order to satisfy the tangent equation we find that when $\mu$ is positive the only allowed values for  $\nu$ are those that are negative.  Consequently, from the cotangent equation we then find that the only allowed values for the imaginary component $k_I$ of $k$ are negative. While we have only shown this to be the case in the numerical examples presented here, in the appendix we show analytically that the square well only admits of  outgoing negative $k_I$ solutions.  For outgoing modes $k_R$ is positive. Thus from (\ref{3.16e}) below, viz. $\hbar^2k_Rk_I/m=-\Gamma_2$, we see that outgoing $PT$ modes only admit of decay, with there being no outgoing modes that grow in time. Consequently, from (\ref{3.14c}),  viz. $\Gamma_2=2\mu\nu$, we see that  $\mu$ and $\nu$ must have opposite signs, just as exhibited above in our numerical evaluation of $T(\mu,\nu,V_0)$ in a few cases, where the only vanishing $T(\mu,\nu,V_0)$ cases are those in which $\mu$ and $\nu$ do have opposite signs. Now the outgoing  boundary conditions given in (\ref{3.10c}) are invariant under $K\rightarrow -K$. This is because the $R<a$ wave function is of the form $A\sin(Kr)/r$, so if we replace $K$ by $-K$ we can replace $A$ by $-A$ and this does not affect the boundary conditions given in (\ref{3.10c}) since they are independent of $A$. Since $K=\gamma(\mu+i\nu)$, the equations given in (\ref{3.11c}) are invariant under $\mu+i\nu \rightarrow -\mu-i\nu$, and there is thus no absolute meaning to the sign of $K$, i.e., no absolute  meaning to the overall signs of $\mu$ and $\nu$, only their relative sign is relevant. In $T(\mu,\nu,V_0)=-T(-\mu,-\nu,V_0)$ and $COT(\mu,\nu,V_0)=COT(-\mu,-\nu,V_0)$,  all that matters is that $\mu$ and $\nu$ have opposite signs. (If for a given $\mu$ and $\nu$ the quantity $T(\mu,\nu,V_0)$ is zero and the quantity ${\rm Im}[COT(\mu,\nu,V_0)]$ is negative, then $T(-\mu,-\nu,V_0)$ is zero and ${\rm Im}[COT(-\mu,-\nu,V_0])$ is negative.) Thus in the outgoing sector we shall take $\mu$ and thus $K_R$ to be positive, and $\nu$ and thus $K_I$ to be negative. 

As well as the lowest lying positive $\mu$ solution there are also higher $\mu$ solutions. We just list one with $V_0=4$. 

\smallskip

For $V_0=4$ we find that $\mu=20.272$, $\nu=\pm 3.018$, $E_2=401.841$, $\Gamma_2=\mp 122.357$.

For $V_0=4$ we find that $T(20.272,+ 3.018,4)=0.003+2.009 i$, $T(20.272,- 3.018,4)=0$.

For $V_0=4$ we find that $COT(20.272,+ 3.018,4)=-20.175-3.032i$, $COT(20.272,- 3.018,4)=20.175-3.032i$,.

\smallskip

In this solution the value found for $\mu$ lies in the $6\pi \leq \mu \leq 13\pi/2$ range, just  as required in order to provide a non-trivial solution to (\ref{3.13c}). With $\nu<0$ we find that $E_2=401.841$ is above threshold, that $\Gamma_2=122.357$ is positive, viz. decay, and that $K_R=20.272$, $K_I=-3.018$, $k_R=20.175$, $k_I=-3.032$, with $k_I$ again being negative, just as required. 

Before comparing these $\mu,\nu$ solutions with Breit-Wigner modes it is instructive to compare them with the $\tan\delta$ solutions as calculated from (\ref{3.2}) and shown in Fig. \ref{tandelta}. For $V_0=4$ we showed that there was only one $\tan\delta$ solution. However, we already presented two $\mu,\nu$ solutions with this same $V_0$.  Further solutions are at $\mu=39.176$, $\nu=-3.672$, and so forth, with (\ref{3.13c}) admitting of an infinite number of solutions. Thus the number of solutions for a given $V_0$ serves as a diagnostic to distinguish $\mu,\nu$ solutions from continuum $\tan\delta$ solutions. Thus continuum resonances and $PT$ resonances can be widely spaced (for the $V_0=4$ case for instance the continuum $\tan\delta$ is infinite solely at $E=5.117$, while the lowest lying  $E_2-i\Gamma_2$ mode is at a quite distant $E=53.295$), and their contributions to the scattering amplitude can be isolated one from the other, though when close they both would have to be included. (In principle, away from continuum resonances we should also include the non-resonant continuum background, but since it is non-resonant at those energies at which the $E_2-i\Gamma_2$ modes  are resonant it is not significant.)

In regard to the Breit-Wigner modes, we additionally note that the fact that all outgoing $PT$ modes only decay and never grow in time  stands in contrast to our above study of Breit-Wigner modes as the Breit-Wigner modes  found above all have  negative $\Gamma_1$. With $\mu$ being greater than $\nu$ and with $\mu^2-\nu^2$ being greater than $V_0$, from (\ref{3.14c}) we see that $E_2$ is above threshold. We thus establish that the square well admits of exact energy eigenvalues of the decaying form $E_2-i\Gamma_2$. The existence of any such exact solutions to the above-threshold Schr\"odinger equation does not appear to have been considered in discussions of scattering off a square well, and constitutes a main result of our study. Also we note that with (\ref{3.11c}) we are being forced into the complex energy plane above threshold. While this is of course standard for the scattering amplitude $f(E)$, we see here that the standard analytic structure of the scattering amplitude originates in the structure of solutions to  the Schr\"odinger equation.

If we now continue from the energy eigenvalue $E_2-i\Gamma_2$  to real $E$ we could only recover the continuum Breit-Wigner solutions in which $\tan \delta_{\rm BW}=\infty$ if according to (\ref{3.4b}) the real parts  of the momenta are such that $a(K,V_0)=(K^2-V_0)^{1/2}\tan K\tan((K^2-V_0)^{1/2})/K+1$ is equal to zero. Evaluating this quantity in the three $V_0=4$, $V_0=8$ and $V_0=16$  cases respectively give values $a(7.582,4)=6.763$,  $a(7.622,8)=5.004$, and $a(7.661,16)=2.117$, with none of them being zero. Even though we do not recover the continuum Breit-Wigner form, in (\ref{3.15d}) below we show that we still get real energy resonances, not at $E=E_1$ but at $E=E_2$. 

If we instead continue in the opposite direction from the continuum Breit-Wigner case to the complex energy eigenvalue case, the continuum Breit-Wigner solutions could only be eigensolutions if their real $K$ as obtained from (\ref{3.4b}) can be identified with a $\mu$ parameter for which $b(\mu,V_0)=\cosh[(\mu^2\cot^2\mu-V_0\cos^2\mu)^{1/2}]-\mu/V_0^{1/2}\sin \mu$ is zero. Evaluating in the three $V_0=4$, $V_0=8$ and $V_0=16$ cases with $K=2.262$, $K=3.266$, $K=6.045$ respectively  give $b(2.262,4)=0.627$, $b(3.266,8)=9.456\times 10^{10}$, and $b(6.045,16)=2.395\times 10^{10}$, again with none of them being zero. (The $\cosh$ factor can be very big when its argument is not at a solution.) Thus the Breit-Wigner $E_1-i\Gamma_1$ modes are not energy eigenvalues, which is perhaps just as well since, as noted above, the square well admits of solutions with $\Gamma_1$ negative, leading to poles that would lie above the real $E$ axis in the complex $E$ plane and be associated with exponential growth in time.

The reason for this continuation mismatch is that the Breit-Wigner modes are determined from real energy solutions to (\ref{3.4b}) where the continuum determined $\tan\delta$ is infinite, while the complex solutions that obey (\ref{3.12c}) are determined by $\tan\delta=-i$. Neither set of solutions could be continued into the other since we do not go through eigensolutions on the way. Thus the correct solutions to use are the $E_2-i\Gamma_2$ solutions. For such solutions we note that if we set  $\tan\delta_{PT}=\Gamma_2/(E_2-E)$ we obtain a scattering amplitude of the form
\begin{align}
f_{PT}(E)=\frac{\Gamma_2}{E_2-i\Gamma_2-E}, \qquad \vert f_{PT}(E)\vert^2=\frac{\Gamma_2^2}{(E_2-E)^2+\Gamma_2^2}.
\label{3.15d}
\end{align}
We thus see that the form $\tan\delta_{PT}=\Gamma_2/(E_2-E)$ does indeed lead to a $\tan\delta_{PT}=-i$ pole in $f_{PT}(E)$ at $E=E_2-i\Gamma_2$, while the continuation of $f_{PT}(E)$ to real $E$ provides  an acceptable alternative to the standard Breit-Wigner $E_1-i\Gamma_1$ discussion, as it will still give  peaks in the  cross-section at $E=E_2$. Since the Breit-Wigner modes are not eigenstates of the scattering Hamiltonian  we could not identify them with physical particles. The complex eigenmodes that obey (\ref{3.12c}) can be associated with physical particles, as manifested in the fact that each $E_2-i\Gamma_2$ pole contribution to the scattering amplitude continues into  a real energy resonance. 

We thus conclude that  the Breit-Wigner $E_1-i\Gamma_1$  modes are not in fact energy eigenstates of the scattering Hamiltonian, and their presence as poles in the scattering amplitude is just an artifact of analytic continuation. This is not fully appreciated in studies of processes such as real energy hadron resonance production since the scattering Hamiltonian is not known and there is no way to check whether or not any resonance that is found might be associated with an eigenstate of the scattering Hamiltonian. However, the complex $E_2-i\Gamma_2$ modes that obey (\ref{3.12c}) can be identified with physical particles, and  each $E_2-i\Gamma_2$ continues into  a real energy resonance. Thus  in the end we can identify real energy cross section resonances with real physical particles, just not as continuum Breit-Wigner ones, but as $PT$ symmetry  ones instead.

\subsection{Radial dependence}

For a state with complex energy $E_2-i\Gamma_2$ where $E_2$ and $\Gamma_2$ are both positive, the wave function time dependence is of the form $\psi(t)=e^{-i(E_2-i\Gamma_2)t/\hbar}$, to thus be an eigenstate of $i\hbar \partial_t$ even though it is  not stationary ($\psi^*(t)\psi(t)$ is not time independent). To specify the spatial behavior, we  set $K=K_R+iK_I$ and $k=k_R+ik_I$, so that  
with $(\hbar^2/2m)K^2=E_2-i\Gamma_2$, $(\hbar^2/2m)k^2=E_2-i\Gamma_2-V_0$, we obtain 
\begin{align}
\frac{\hbar^2}{2m}(K_R^2-K_I^2)=E_2,\qquad \frac{\hbar^2K_RK_I}{m}=-\Gamma_2,\qquad \frac{\hbar^2}{2m}(k_R^2-k_I^2)=E_2-V_0,\qquad \frac{\hbar^2k_Rk_I}{m}=-\Gamma_2.
\label{3.16e}
\end{align}
Taking $K_R$ and $k_R$ to be positive, then with positive $E_2$ and $\Gamma_2$  we see that $K_I$ and $k_I$ are both negative.

Thus  on setting $C=-i\cos\delta B$ (as per (\ref{3.9c})) the outgoing $PT$ wave function is given by the outgoing 
\begin{align}
\psi_{PT}(r<a,t)=A\frac{\sin((K_R+iK_I)r)}{r}e^{-iE_2t-\Gamma_2 t}, \qquad \psi_{PT}(r>a,t)=C\frac{e^{ik_Rr- k_Ir}}{r}e^{-iE_2t- \Gamma_2 t}.
\label{3.17e}
\end{align}
With the non-oscillating part of the $\psi_{PT}(r>a,t)$ wave function behaving as $e^{-k_Ir-\Gamma_2t}$ with $k_I<0$ and $\Gamma_2>0$, $\psi_{PT}(r>a,t)$ is known as a Gamow vector. With its exponential growth in $r$ ($k_I$ being negative) the time decaying $\psi_{PT}(r>a,t)$ has a Dirac inner product that is not square-integrable. The Hamiltonian is self-adjoint when it acts in the bound state sector but is not self-adjoint when it acts on $\psi_{PT}(r>a,t)$.  However, with complex energy eigenvalues the Hamiltonian is not Hermitian, but as we explain in Sec. \ref{S4} it is instead pseudo-Hermitian, and this will lead  to a different inner product, one that will enable us to address the lack of square integrability.

With parity having no effect on  $\psi_{PT}(r,t)$ since $r=(x^2+y^2+z^2)^{1/2}$  is not affected by parity, under either $PT$ or time reversal $\psi_{PT}(r,t)$ conjugates into the ingoing 
\begin{align}
\psi^{(c)}_{PT}(r<a,t)=A^*\frac{\sin((K_R-iK_I)r)}{r}e^{-iE_2t+\Gamma_2 t}, \qquad \psi^{(c)}_{PT}(r>a,t)=C^*\frac{e^{-ik_Rr- k_Ir}}{r}e^{-iE_2t+ \Gamma_2 t},
\label{3.18d}
\end{align}
to give a mode with energy $E_2+i\Gamma_2$, with $\psi^{(c)}(r,t)$ also being an exact eigenstate of the square well Hamiltonian.  
The key feature  of $\psi^{(c)}_{PT}(r>a,t)$ for our purposes here is that just like $\psi_{PT}(r>a,t)$ it diverges exponentially in space (under $PT$ we have $t\rightarrow -t$ and $r\rightarrow r$), something we demonstrate analytically for the arbitrary $\psi^{(c)}_{PT}(r>a,t)$ in the appendix. Thus we could not use $\psi^{(c)}_{PT}(r>a,t)$  to form a spatial convergent matrix element  with $\psi_{PT}(r>a,t)$. To accommodate the exponential growth of $\psi_{PT}(r>a,t)$ we will have to look beyond the square well itself. This we will do below in Sec. \ref{S5}. In fact we will need to do so anyway since outgoing waves that are produced in  scattering experiments are not growing exponentially in space. 

That we do find complex energy eigenvalues is initially a little puzzling since the square well potential and the associated secular equation  $\vert H-\lambda I\vert=0$ that determines energy eigenvalues are both real. However, as well as real solutions a real secular equation can also have solutions that are complex provided that they appear in complex conjugate pairs, just as we show below has to anyway be the case due to the $PT$ symmetry that the square well Hamiltonian has. Given the $E_2-i\Gamma_2$ solution, the second solution (the time reversed one) thus has to have energy eigenvalue $E_2+i\Gamma_2$.  From (\ref{3.17e}) it follows that since $k_R$ is positive the $r>a$ radial function associated with $E_2-i\Gamma_2$ is of the outgoing form $e^{ik_Rr-k_Ir}$ at $t=\infty$. Under time reversal (or $PT$) it conjugates into the ingoing $e^{-ik_Rr-k_Ir}$ at $t=-\infty$. With $E_2+i\Gamma_2$ thus being associated with an ingoing mode it is not relevant in a scattering experiment. However,  in regard to the outgoing $E_2-i\Gamma_2$ mode, we note that while it decays exponentially in time it grows exponentially in space ($k_I$ being negative), to thus be a Gamow vector. We will discuss below how to accommodate this exponential spatial growth, and note that we would find the same radial behavior for the  Breit-Wigner $E_1-i\Gamma_1$ mode as its  momentum is also continued into the complex plane, just as Gamow had noted.

\section {$PT$ symmetry and pseudo-Hermiticity}
\label{S4}

The existence of complex conjugate pairs of energy eigenvalues is a characteristic of the antilinear $PT$ symmetry program of Bender and collaborators (see e.g. the typical \cite{Bender1998,Bender1999,Mostafazadeh2002,Bender2002,Bender2007,Makris2008,Bender2008a,Bender2008b,Guo2009,Bender2010,Special2012,Theme2013,ElGanainy2018,Bender2018,Fring2021,Mannheim2018a,Mannheim2025}), a program that has now realized of order 16,000 peer-reviewed papers.  The notion of antilinear symmetry dates back to Wigner's study of time reversal \cite{Wigner1960}, and as developed through the related pseudo-Hermiticity that $PT$ symmetry entails enables one to construct a 
time-independent  probability amplitude for the eigenstates of complex conjugate energy pairs. The essential feature of the program for our purposes here is that the Schr\"odinger equation $H\vert\psi\rangle=E\vert \psi\rangle$ only specifies the ket, leaving the bra free. The bra does not need to be the Hermitian conjugate of the ket. It can instead be the $PT$ conjugate of the ket (or $CPT$ conjugate in the relativistic case \cite{Mannheim2018a}, where $C$ is charge conjugation). Equivalently, the bra can be the left-eigenvector of the Hamiltonian rather than the Hermitian conjugate of the right-eigenvector. The $PT$ symmetry program thus recognizes that quantum theory allows the dual space to  be more general than the standard Hermitian conjugate one. This allows us to make sense of Hamiltonians that are not Hermitian, such as the square well  Hamiltonian in its complex energy eigenvalue realization.

To be specific, if  we introduce some general antilinear operator $\hat{A}$ that commutes with a Hamiltonian $\hat{H}$, and consider eigenvalues $E$ and eigenfunctions $e^{-iEt}\vert \phi\rangle$ of $\hat{H}$ that obey $\hat{H}\vert \phi \rangle=E\vert \phi\rangle$ we obtain
\begin{align}
\hat{H}\hat{A}\vert \phi\rangle=\hat{A}\hat{H}\vert \phi\rangle=\hat{A}E\vert \phi\rangle=E^*\hat{A}\vert \phi\rangle.
\label{4.1}
\end{align}
Thus for every eigenvalue $E$ with eigenvector $\vert \phi\rangle$ there is an eigenvalue $E^*$ with eigenvector $\hat{A}\vert \phi\rangle$. Thus as first noted by Wigner in his study of time reversal invariance, antilinear symmetry implies that energy eigenvalues are either real or in complex conjugate pairs. For our  purposes here where $\hat{A}=PT$,  we  note that with the wave function $\psi_{PT}(r,t)$ given in (\ref{3.17e}) having eigenvalue $E_2-i\Gamma_2$, according to (\ref{4.1}) the state with eigenvalue $E_2+i\Gamma_2$  must be given by the $PT$ conjugate of $\psi_{PT}(r,t)$, which is exactly how we constructed $\psi^{(c)}_{PT}(r,t)=PT\psi_{PT}(r,t)$ in (\ref{3.18d}).  Now while we used the secular equation  $\vert H-\lambda I\vert=0$ in the above in order to establish that there had  to be complex conjugate pairs of energy eigenvalues, we need (\ref{4.1}) to establish that the associated radial wave functions are also in a $PT$ conjugate pair ($\vert \phi\rangle$ and $\hat{A}\vert \phi\rangle$).

To obtain time independence and probability conservation in a non-Hermitian case we introduce a time-independent operator $\hat{V}$, and with $i\partial_t\vert \psi\rangle=\hat{H}\vert \psi\rangle$, $-i\partial_t\langle \psi \vert=\langle \psi \vert \hat{H}^{\dagger}$, we evaluate
\begin{align}
i\frac{\partial}{\partial t}\langle \psi(t)\vert \hat{V}\vert \psi(t)\rangle=\langle \psi(t)\vert (\hat{V}\hat{H}-\hat{H}^{\dagger}\hat{V})\vert \psi(t)\rangle.
\label{4.2}
\end{align}
Thus $\langle \psi(t)\vert \hat{V}\vert \psi(t)\rangle$ will be time independent and probability will be conserved if
there exists a $\hat{V}$ that obeys 
\begin{align}
\hat{V}\hat{H}=\hat{H}^{\dagger}\hat{V},\quad \hat{V}\hat{H }\hat{V}^{-1}=\hat{H}^{\dagger},
\label{4.3}
\end{align}
with the second condition requiring that $\hat{V}$ be invertible, something we take to be the case here. The $\hat{V}\hat{H}\hat{V}^{-1}=\hat{H}^{\dagger}$ condition is  known as pseudo-Hermiticity and implements probability conservation. Pseudo-Hermiticity was introduced by Dirac \cite{Dirac1942} and Pauli \cite{Pauli1943} in their study of indefinite metric quantum field theories (a recent discussion of which may be found in \cite{Mannheim2024}), and was connected to $PT$ symmetry in \cite{Mostafazadeh2002}. Pseudo-Hermiticity enables us to construct a time-independent inner product when the standard  $\langle \psi(t)\vert \psi(t)\rangle$ is not time independent.

The $\hat{V}\hat{H}\hat{V}^{-1}=\hat{H}^{\dagger}$ condition entails that the relation between $\hat{H}$ and $\hat{H}^{\dagger}$  is isospectral. Thus every eigenvalue of $\hat{H}$ is an eigenvalue of $\hat{H}^{\dagger}$. Consequently the eigenvalues of $\hat{H}$ are  either real or in complex conjugate pairs. But this is the antilinear symmetry condition. Thus for any $\hat{H}$ whose eigenspectrum is complete there always will be a $\hat{V}$ operator if $\hat{H}$ has an antilinear symmetry \cite{Mannheim2018a}, with $\langle \psi(t)\vert \hat{V}\vert \psi(t)\rangle$ being the most general inner product that one could introduce that is probability conserving. If we introduce right-eigenvectors $\vert R\rangle$ of $\hat{H}$ according to  $\hat{H}\vert R\rangle=E\vert R \rangle$, we have
\begin{align}
\langle R\vert \hat{H}^{\dagger}=E^*\langle R\vert=\langle R\vert \hat{V}\hat{H }\hat{V}^{-1},\qquad \langle R\vert \hat{V}\hat{H }=E^*\langle R\vert \hat{V}.
\label{4.4}
\end{align}
Consequently, we  can identify a left eigenvector $\langle L\vert=\langle R\vert \hat{V}$, and can write the inner product as $\langle R\vert \hat{V}\vert R\rangle=\langle L\vert R\rangle$. Thus in general we can  identify the left-right inner product as the most general probability-conserving inner product in the antilinear case, a form that could perhaps be  anticipated since a Hamiltonian cannot have any more eigenvectors than its left and right ones. 

In the above we had noted that in the particular case when $V_0=\hbar^2\pi^2/8ma^2$ there is a bound state when $E$ is slightly less than $V_0$, a continuum resonance when $E$ is slightly larger than $V_0$, and an exceptional point at $E=V_0$. Such a pattern is characteristic oi $PT$ symmetry (see e.g. \cite{Mannheim2018a}), with the transition as we change the energy from  real energy eigenvalues to complex ones going through an exceptional point.

\subsection{Application to complex conjugate pairs of eigenstates}

To illustrate the role of $\hat{V}$ we combine the two complex conjugate eigenvalues $E_2\pm i\Gamma_2$ of any pair we might be interested in (not just a square well pair) into a diagonal  two-dimensional matrix of the form 
\begin{align}
M_2=\begin{pmatrix}E_2+i\Gamma_2&0\\ 0&E_2-i\Gamma_2 \end{pmatrix}.
\label{4.5}
\end{align}
As constructed, the matrix $M_2$  is $PT$ symmetric under $P=\sigma_1$, $T=K$, where $K$ denotes complex conjugation, to thus naturally have complex conjugate eigenvalues. The operator  $V=i\sigma_2$ effects $VM_2V^{-1}=M_2^{\dagger}$. With this $V$  the eigenkets, eigenbras, and the  orthogonality and closure relations associated with  $M_2$ are given as in  \cite{Mannheim2018a} as
\begin{align}
&u_+=e^{-iE_2t+\Gamma_2 t}\begin{pmatrix}1\\ 0\end{pmatrix}, \qquad u_-=e^{-iE_2t-\Gamma_2 t}\begin{pmatrix}0\\ 1\end{pmatrix}, \qquad u_+^{\dagger}V=e^{iE_2t+\Gamma_2t}\begin{pmatrix}0&1 \end{pmatrix},\qquad u^{\dagger}_-V=e^{iE_2t-\Gamma_2t}\begin{pmatrix}-1 &0\end{pmatrix},
\nonumber\\
&u_{\pm}^{\dagger}Vu_{\pm}=0,\qquad u_{-}^{\dagger}Vu_{+}= -1,\qquad u_{+}^{\dagger}Vu_{-}=1,
\qquad u_{-}u^{\dagger}_{+}V-u_{+}u^{\dagger}_{-}V=I.
\label{4.6}
\end{align}
The appearance of the $-1$ factor in $u_{-}^{\dagger}Vu_{+}$ is not indicative of any possible negative norm ghost problem since  $u_{-}^{\dagger}Vu_{+}$ is a transition matrix element between two different states (with energies $E_2\pm i\Gamma_2$), and not the overlap of a state with its own conjugate.
As we see, all of the $V$-based inner products are time independent, with the only non-vanishing ones being the ones that link the decaying and growing modes. As the population of one level decreases the population of the other level increases by the same amount, with probability conservation requiring that we consider both levels together and not just one (the decaying one) as is usually done in the  standard Breit-Wigner approach.  

For the specific square well case, on multiplying the wave functions in (\ref{4.6}) by the spatial terms given in (\ref{3.17e}) and (\ref{3.18d}) and identifying $\psi_{PT}$ with $u_-$ and $\psi^{(c)}_{PT}$ with $u_+$ then leads to 
\begin{align}
&\psi^{\dagger}_{PT}(r<a,t)V\psi_{PT}(r<a,t)=0,\quad  \psi^{(c)\dagger}_{PT}(r<a,t)V\psi^{(c)}_{PT}(r<a,t)=0, 
\nonumber\\
& \psi^{\dagger}_{PT}(r>a,t)V\psi_{PT}(r>a,t)=0,\quad \psi^{(c)\dagger}_{PT}(r>a,t)V\psi^{(c)}_{PT}(r>a,t)=0,
\nonumber\\
&\psi^{\dagger}_{PT}(r<a,t)V\psi^{(c)}_{PT}(r<a,t)=-A^*A\frac{\sin((K_R- iK_I)r)\sin((K_R- iK_I)r)}{r^2},
\nonumber\\
&\psi^{\dagger}_{PT}(r>a,t)V\psi^{(c)}_{PT}(r>a,t)= -C^*C\frac{e^{-2k_Ir}}{r^2},
\nonumber\\
&\psi^{(c)\dagger}_{PT}(r<a,t)V\psi_{PT}(r<a,t)=A^*A\frac{\sin((K_R+iK_I)r)\sin((K_R+ iK_I)r)}{r^2},
\nonumber\\
& \psi^{(c)\dagger}_{PT}(r>a,t)V\psi_{PT}(r>a,t)= C^*C\frac{e^{-2k_Ir}}{r^2},
\label{4.7}
\end{align}
with all exponentially growing and exponentially falling terms in time cancelling each other. However the exponentially growing $e^{-2k_Ir}$ term is not cancelled. We will address this issue below.

\subsection{The propagator}

We can discuss the square well eigenstates without regard to any scattering experiment, a situation  that would only involve the outgoing $\psi_{PT}$ and not the ingoing $\psi^{(c)}_{PT}$. To discuss the associated propagator  we introduce the resolvent $1/(E-H)$, though in a scattering experiment the resolvent  would be projected onto  $\psi_{PT}$, with the time-reversed scattering being projected onto $\psi^{(c)}_{PT}$. However, for the full propagator  both states are involved.  Then from (\ref{4.6}) we obtain the associated propagator, and in analog to the discussion in   \cite{Mannheim2018a,Mannheim2013} obtain 
\begin{align}
D_{PT}(E)=\frac{u_{+}^{\dagger}Vu_{-}}{E-(E_2-i\Gamma_2)}+\frac{u_{-}^{\dagger}Vu_{+}}{E-(E_2+i\Gamma_2)}=\frac{1}{E-E_2+i\Gamma_2}-\frac{1}{E-E_2-i\Gamma_2}=-\frac{2i\Gamma_2}{(E_2-E)^2+\Gamma_2^2}.
\label{4.8}
\end{align}
We thus obtain the same negative imaginary sign for the propagator as obtained in the standard (\ref{2.1}) that only contained one pole, so that operationally (\ref{4.8}) and (\ref{2.1})  are equivalent. It is this equivalence that  enabled Lee and Wick \cite{Lee1969} to consider the relativistic generalization of (\ref{4.8}), viz. a propagator of the form $1/(k^2-M^2+iN^2)-1/(k^2-M^2-iN^2)$, in order to obtain better large $k^2$ behavior (viz. $1/k^4$) than a standard $1/k^2$ propagator, to  then control quantum field theory renormalization without jeopardizing unitarity \cite{footnote6}.

\section{Treating the exponentially growing spatial  terms}
\label{S5}
\subsection{Rigged Hilbert space approach}

With the radial wave functions of both $\psi_{PT}(r,t)$ and $\psi^{(c)}_{PT}(r,t)$  growing exponentially as $e^{-k_Ir}/r $ ($k_I$ being negative) the standard treatment of such  states (known as  Gamow vectors) is to place them in a dual space $\Phi^*$ of a rigged Hilbert  space (a nested Gelfand triple $\Phi \subset{\mathcal H}\subset \Phi^*$ of Hilbert spaces), with a sufficiently convergent set of test functions in the test function space $\Phi$ so that the overlap of states in $\Phi^*$ and $\Phi$ yields a finite probability (see e.g. \cite{Bohm1989}).   The Hilbert space $\mathcal{H}$ consists of the square-integrable modes of the Hamiltonian $H$, and for the square well that would be the bound states. The test functions considered in the Hilbert space $\Phi$ would have to not just be square integrable, they would have to yield well-behaved overlaps with the non-square-integrable states in $\Phi^{*}$. Thus for the $\psi_{PT}(r,t)$ and $\psi^{(c)}_{PT}(r,t)$ states in $\Phi^*$, viz. for 
\begin{align}
\psi_{PT}(r<a,t)&=A\frac{\sin((K_R+iK_I)r)}{r}e^{-iE_2t-\Gamma_2 t}, \qquad \psi_{PT}(r>a,t)=C\frac{e^{ik_Rr- k_Ir}}{r}e^{-iE_2t- \Gamma_2 t},
\nonumber\\
\psi^{(c)}_{PT}(r<a,t)&=A^*\frac{\sin((K_R-iK_I)r)}{r}e^{-iE_2t+\Gamma_2 t}, \qquad \psi^{(c)}_{PT}(r>a,t)=C^*\frac{e^{-ik_Rr- k_Ir}}{r}e^{-iE_2t+ \Gamma_2 t},
\label{5.1a}
\end{align}
we would need states in $\Phi$ that decay exponentially in space, and  to counterbalance the exponential time behavior they would need to have compensating time behaviors. Thus with $K_R>0$, $K_I<0$, $k_R>0$, $k_I<0$, $E_2>0$, $\Gamma_2>0$, for $\psi_{PT}(r,t)$ and $\psi^{(c)}_{PT}(r,t)$ we introduce the following square-integrable wave functions in the Hilbert space $\Phi$: 
\begin{align}
\psi_{AG}(r<a,t)&=D\frac{\sin((K_R+iK_I)r)}{r}e^{-iE_2t+\Gamma_2 t}, \qquad \psi_{AG}(r>a,t)=F\frac{e^{ik_Rr+k_Ir}}{r}e^{-iE_2t+ \Gamma_2 t},
\nonumber\\
\psi^{(c)}_{AG}(r<a,t)&=D^*\frac{\sin((K_R-iK_I)r)}{r}e^{-iE_2t-\Gamma_2 t}, \qquad \psi^{(c)}_{AG}(r>a,t)=F^*\frac{e^{-ik_Rr+k_Ir}}{r}e^{-iE_2t- \Gamma_2 t}.
\label{5.2a}
\end{align}
The outgoing $k_R>0$ $\psi_{AG}(r>a,t)$ grows exponentially in time while falling exponentially in space, to thus complement the outgoing Gamow vector $\psi_{PT}(r>a,t)$ that grows exponentially in space while falling exponentially in time. (With $\hbar^2k_Rk_I/m=-\Gamma_2$ and $k_R$ positive, $k_I$ and $\Gamma_2$ have opposite signs, so that negative $\Gamma_2$ is accompanied by positive $k_I$, i.e., exponential growth in time and exponential fall in space.) We shall refer to $\psi_{AG}(r>a,t)$ as an anti-Gamow ($AG$) vector. Its $PT$ conjugate $\psi^{c}_{AG}(r>a,t)$ falls exponentially in time and falls  exponentially in space, to thus complement  the $PT$ conjugate wave function $\psi^{(c)}_{PT}(r>a,t)$ that grows exponentially in space and grows exponentially in time. The  $\psi_{PT}(r,t)$ and  $\psi^{(c)}_{PT}(r,t)$ wave functions are non-square-integrable eigenstates of the square well $H$, while  $\psi_{AG}(r,t)$ and  $\psi^{(c)}_{AG}(r,t)$ are square integrable but are not eigenstates of the square well $H$ (they do not satisfy the boundary conditions in  (\ref{3.1})).

In the overlap combinations
\begin{align}
\psi^{\dagger}_{AG}(r>a,t)\psi_{PT}(r>a,t)=\frac{F^*C}{r^2},\qquad \psi^{(c)\dagger}_{AG}(r>a,t)\psi^{(c)}_{PT}(r>a,t)=\frac{FC^*}{r^2},
\label{5.3a}
\end{align}
all exponential terms cancel, with the overlaps behaving as $1/r^2$, viz.  just like the fully acceptable continuum spherical mode overlaps. However, while all spatial exponentials between $\psi_{PT}(r>a,t)$ and $\psi^{(c)_\dagger}_{AG}(r>a,t)$ or between $\psi^{(c)}_{PT}(r>a,t)$ and $\psi^{\dagger}_{AG}(r>a,t)$ would cancel, exponential time dependence would not cancel. Now we had noted above in (\ref{4.3}) that that with pseudo-Hermiticity we can construct time independent matrix elements. Thus in analog to (\ref{4.5}) we  introduce the four-dimensional matrix $M_4$ and the $V_4$ matrix that effects $V_4M_4V_4^{-1}=M_4^{\dagger}$:
\begin{align}
M_4=\begin{pmatrix}E_2+i\Gamma_2&0&0&0\\ 0&E_2-i\Gamma_2 &0&0\\
 0&0&E_2+i\Gamma_2&0\\ 0&0&0&E_2-i\Gamma_2\end{pmatrix},\qquad V_4=\begin{pmatrix}0&1&0&0\\ -1&0&0&0\\
 0&0&0&1\\ 0&0&-1&0\end{pmatrix}.
\label{5.4a}
\end{align}
We associate the upper $2\times 2$ block with $\psi_{AG}(r,t)$ and $\psi_{PT}(r,t)$, and associate the lower $2\times 2$ block with $\psi^{(c)}_{PT}(r,t)$ and $\psi^{(c)}_{AG}(r,t)$. Then, since $M_4$ is block diagonal there are no overlaps between two sectors, and now just as (\ref{4.7})  all overlaps are both time independent and spatially independent. As far as we know, this is the first time pseudo-Hermitiicity has been applied to a rigged Hilbert space treatment of states that grow or fall exponentially in time or space. 

\subsection{Decay product approach}

Apart from using a rigged Hilbert space, it is possible to treat exponentially growing or falling modes by taking into consideration the decay products that are actually produced by  time decaying states such as  the outgoing Gamow vector $\psi_{PT}(r,t)$ in a two-stage process in which $\psi_{PT}(r,t)$ is first produced and then decays. Specifically, since the mode $\psi_{PT}(r,t)$ that obeys $E-V_0=(\hbar^2/2m)k^2$ is decaying it must decay into a decay product that grows in time. Loss or gain of energy or momentum in the Gamow mode is balanced by gain or loss in whatever outgoing mode the Gamow mode decays into. Thus such decay product modes must also obey $E-V_0=(\hbar^2/2m)k^2$ with $k_R>0$, and since they behave as $e^{-iE_2t+\Gamma_2t}$, they must behave as $e^{ik_Rr+k_Ir}$ in space, where $\Gamma_2>0$ and $k_I<0$. We recognize such a state as an anti-Gamow state. Thus the Gamow vector that is produced in the first stage via  its scattering by the square well propagates in the second stage into an anti-Gamow vector, and by continuity (no new square well type boundary) the anti-Gamow vector  wave function  must be given by
\begin{align}
\psi_{AG}(r>a,t)=A\frac{e^{ik_Rr+k_Ir}}{r}e^{-iE_2t+ \Gamma_2 t},
\label{5.5a}
\end{align}
with its  overlap with $\psi_{PT}(r>a,t)$ being given by the well behaved 
\begin{align}
\psi^{\dagger}_{AG}(r>a,t)\psi_{PT}(r>a,t)=\frac{A^*A}{r^2},\qquad \psi^{\dagger}_{PT}(r>a,t)\psi_{AG}(r>a,t)=\frac{A^*A}{r^2}.
\label{5.6a}
\end{align}
All exponential terms cancel, with the overlaps behaving as $1/r^2$, viz.  just like the fully acceptable continuum spherical mode overlaps. (For continuum modes $\psi(q)=e^{iqr}/r$, $\psi(q^{\prime})=e^{iq^{\prime}r}/r$ we have $\int d^3r e^{-iq^{\prime}r}e^{iqr}/r^2=4\pi\int_0^{\infty}dr r^2e^{i(q-q^{\prime})r}/r^2=4\pi\int_0^{\infty}dr e^{i(q-q^{\prime})r}$.) Thus using the $\langle L\vert R\rangle$  pseudo-Hermiticity inner product rather than the Dirac $\langle R\vert R\rangle$ enables us to treat the complex conjugate $E_2\pm i\Gamma_2$ modes in exactly the same way as we treat plane waves with real energy. 

Now while $\psi_{AG}(r>a,t)$ is not an eigenstate of the square well Hamiltonian, we note that even though both its energy and momentum are complex, it is an s-wave eigenstate of the Hamiltonian $H_0=-(\hbar^2/2m)\nabla^2+V_0$. Moreover, so is the $\psi_{PT}(r>a,t)$ Gamow vector, with the Gamow vector having energy and momentum $E=E_2-i\Gamma_2$, $k=k_R+ik_I$, and the anti-Gamow vector having energy and momentum $E=E_2+i\Gamma_2$, $k=k_R-ik_I$. $H_0$ thus serves as the Hamiltonian for the second stage of the decay process, with the square well Hamiltonian having served for the first stage. We can discuss the second stage using an $M_2$ that is this  $H_0$, with $u_+$ in (\ref{4.6}) being identified with the anti-Gamow vector, and $u_-$ being identified with the Gamow vector, and $V=i\sigma_2$ effecting a transition between a state with energy and  momentum $E_2- i\Gamma_2$ and $k_R+ ik_I$ and one with energy and momentum  $E_2+ i\Gamma_2$ and $k_R - ik_I$ ($E_2>0$, $k_R>0$, $\Gamma_2>0$ and $k_I<0$ in both cases).
Then all overlaps other than those given in  (\ref{5.6a}) vanish, just as required of pseudo-Hermiticity, with (\ref{5.6a}) being replaced by 
\begin{align}
\psi^{\dagger}_{AG}(r>a,t)V\psi_{PT}(r>a,t)=\frac{A^*A}{r^2},\qquad \psi^{\dagger}_{PT}(r>a,t)V\psi_{AG}(r>a,t)=-\frac{A^*A}{r^2}.
\label{5.7b}
\end{align}
As per (\ref{4.8}) the decay of the Gamow vector can be associated with the propagator
\begin{align}
D_{PT}(E)=\frac{u_{+}^{\dagger}Vu_{-}}{E-(E_2-i\Gamma_2)}+\frac{u_{-}^{\dagger}Vu_{+}}{E-(E_2+i\Gamma_2)}=\frac{1}{E-E_2+i\Gamma_2}-\frac{1}{E-E_2-i\Gamma_2}=-\frac{2i\Gamma_2}{(E_2-E)^2+\Gamma_2^2},
\label{5.8b}
\end{align}
with poles at $E_2-i\Gamma_2$ and $E_2+i\Gamma_2$.

The presence of non-square-integrable or non-stationary modes is not ordinarily associated with a Hamiltonian $H_0$ that is essentially a free system (just the zero of energy has been shifted by a constant), but we for instance note that a free neutron spontaneously decays  into a proton, electron, and an electron anti-neutrino. On its own $\psi_{PT}(r,t)$ would not be associated with a free system. However when accompanied by $\psi_{AG}(r,t)$ it can be, with $H_0$ having a pseudo-Hermitian structure in this sector.

However, there has to be some specific interaction that would spontaneously convert a Gamow vector into an anti-Gamow vector, an interaction that causes one system to decay into another. (For instance, the weak interaction causes $n\rightarrow p+e^-+\bar{\nu}_e$.) In any such process we note that as the population of the decaying state decreases that of the decay product increases, with the total population of the two states remaining constant, viz. probability conservation. All that is needed for our purposes here is that the initial configuration for the second stage (the in state for the second stage) be a Gamow vector that was produced in the first stage, and the final configuration (the out state for the second stage) be an anti-Gamow vector, as that will take care of all exponential fall and growth in both space and time. In the pseudo-Hermitian case the time evolution operator $U=e^{-iHt}$ obeys the pseudo-unitary $UV^{-1}U^{\dagger}V=I$ and the S-matrix which evolves in-states into out-states obeys $SV^{-1}S^{\dagger}V=I$. In terms of the in and out states the S-matrix is given not by $\sum\vert R_{in}\rangle \langle R_{out}\vert$ but by $\sum\vert R_{in}\rangle \langle L_{out}\vert=\sum\vert R_{in}\rangle \langle R_{out}\vert V$ (see e.g. \cite{Mannheim2013}).

With (\ref{5.8b}) there is only one observable resonance (not two)  in a closed system such as an elastic scattering process below the inelastic threshold  since $E_2-i\Gamma_2$ describes the transition from the in-state to the resonance and $E_2+i\Gamma_2$ describes the subsequent transition from the resonance to the out-state. (A typical case is $\pi^++p$ elastic scattering  in which a short-lived, unstable  1232 MeV $\Delta^{++}$ resonance with a 120 MeV width is produced and  then decays into a $\pi^++p$ out-state.) We recognize such processes as the adiabatic switching procedure of Gell-Mann and Low, as generalized here to cases where the free complex energy plane wave  in-states and out-states are not stationary, with the out-states acquiring a phase due to the intervening interaction that caused the transition between the in and out states to take place.

If all we have is a decay (the Breit-Wigner case) then the system is considered to be open (typically occurring when the decaying state is long lived), with the decay products appearing outside the scattering region. However,  with both a Gamow vector and an anti-Gamow vector the system can be a closed one if both modes are produced in the scattering region (short-lived decay case). In that case  there is no need to rig the Hilbert space, with an anti-Gamow mode that is not an eigenstate of a Hamiltonian (the square well Hamiltonian) being replaced by an anti-Gamow mode that is an eigenstate of a Hamiltonian (viz. $H_0$). In the standard discussion of the Breit-Wigner approach to scattering the loss of Hermiticity entailed in having states that behave as $e^{-i(E_1-i\Gamma_1)t}$  is thought to occur because the system is an open system that is embedded in a larger system in such a way that Hermiticity is then thought to be restored. In such a situation the decay products associated with $e^{-iE_1t-\Gamma_1t}$  decay into the larger system. States in the larger system will then grow as $e^{-iE_1t+\Gamma_1t}$. Even if Hermiticity is not restored in the larger system,  probability conservation is restored since as the population of the decaying system decreases that of the growing system increases accordingly. However, with pseudo-Hermiticity probability conservation is achieved within a closed system that contains both the decaying modes and their decay products.

While  $D_{PT}(E)$ of (\ref{5.8b}) gives the energy dependence of the propagator,  how we determine its time dependence, viz. 
\begin{align}
D_{PT}(t)=\frac{1}{2\pi }\int^{\infty}_{-\infty} dEe^{-iEt}D_{PT}(E),
\label{5.9b}
\end{align}
depends on the form of the  contour that we choose in the complex $E$ plane. There are two poles in $D_{PT}(E)$ as given in (\ref{5.8b}), one above the real $E$ axis and one  below, and we can suppress the lower-half circle contribution when $t>0$, and suppress the upper-half circle contribution when $t<0$. However, in order to be able to combine the two propagators in (\ref{5.8b}) so as to give the final form for the propagator that is also given in (\ref{4.8}), just as in \cite{Lee1969}  we need the two poles in $D_{PT}(E)$ to be in the same contour. We thus deform the contour around the upper-half plane pole at $E=E_2+i\Gamma_2$ in (\ref{4.8}) so that it contributes when we close below. And with both poles then being in the same contour, via contour integration we obtain
\begin{align}
D_{PT}(t)=-i\theta(t)[e^{-iE_2t-\Gamma_2 t}-e^{-iE_2t+\Gamma_2 t}].
\label{5.10b}
\end{align}
Thus we obtain both a forward in time growing mode and a forward in time decaying mode, with the time advanced mode not growing backward in time or being acausal, which it would have been had we located the $E=E_2+i\Gamma_2$ pole in an upper-half plane contour. 

Thus, as per \cite{Mannheim2018a},  we obtain a time-advanced, negative-width $\Delta T=-\hbar/\Gamma_2$  generalization of the standard positive-width time delay $\Delta T=\hbar/\Gamma_2$ produced by a potential well. And not only that, being equal and opposite in sign the time advance and time delay  cancel each other identically. This results in no net  time advance or time delay  in processes such as atomic energy level excitation and decay even though the levels have linewidths. Consequently, the  decay time of an excited atomic level is not equal to $\hbar$ divided by the linewidth of the level, but is instead an essentially instantaneous decay with a time delay given by $\hbar$ divided by the energy difference between the level and the level it  decays into. As  discussed in \cite{Mannheim2025} we note that in experiments with ultracold rubidium atoms, an effect of this type   has been reported, with  the authors of \cite{Sinclair2022} noting that photons are scattered long before the atom’s spontaneous lifetime has elapsed, while the authors of   \cite{Angulo2024}  report detecting a negative time delay.

We can associate the time advance  and time delay with energy-dependent phase shifts of the form
\begin{align}
\tan\delta_-(E)=\frac{\Gamma_2}{E_2-E},\qquad \tan\delta_+(E)=-\frac{\Gamma_2}{E_2-E}.
\label{5.11b}
\end{align}
At the poles they obey $\tan\delta_-(E_2-i\Gamma_2)=-i$, $\tan\delta_+(E_2+i\Gamma_2)=-i$, just as outgoing mode poles should. (As noted above in our discussion of (\ref{3.8c}), outgoing modes such as the Gamow and anti-Gamow modes of interest to us here all have $\tan\delta =-i$.) At real $E=E_2$ they obey $\delta_-(E_2)=\pi/2$, $\delta_+(E_2)=-\pi/2$, just as resonances should, while respectively giving the time delay and time advance described above. Together, they give a scattering amplitude
\begin{align}
&f_{PT}(E)=\frac{\tan\delta_-(E)}{1-i\tan\delta_-(E)}+\frac{\tan\delta_+(E)}{1-i\tan\delta_+(E)}=\frac{\Gamma_2}{E_2-i\Gamma_2-E}-\frac{\Gamma_2}{E_2+i\Gamma_2-E}=\frac{2i\Gamma_2^2}{(E_2-E)^2+\Gamma_2^2},
\nonumber\\
&\vert f_{PT}(E)\vert^2=\frac{4\Gamma_2^4}{[(E_2-E)^2+\Gamma_2^2)]^2},\qquad \vert f_{PT}(E_2)\vert^2=4.
\label{5.12b}
\end{align}
Despite the presence of two poles in the scattering amplitude, in the $\vert f_{PT}(E)\vert^2$ cross-section there is only one resonance peak for each $E_2$ with neither $f_{PT}(E)$ nor $\vert f_{PT}(E)\vert^2$ being sensitive to the sign of $\Gamma_2$, even as $f_{\rm BW}(E)$ is sensitive to the sign of $\Gamma_1$. Also $f_{PT}(E)$ is pure imaginary. However, $f_{\rm BW}(E)$ is not, though its real part vanishes at resonance.

We recognize (\ref{5.12b}) as being of the same form as $D_{PT}(E)$.  Also we note that since $\delta_+(E)=-\delta_-(E)$, it follows that $f_{PT}(E)=e^{i\delta_-(E)}\sin\delta_-(E)-e^{-i\delta_-(E)}\sin\delta_-(E)=2i\sin^2\delta_-(E)$, to thus give a real energy resonance at $\delta_-(E)=\pi/2$, viz. $E=E_2$. The presence of resonances in the real energy scattering cross-section thus indicates that the Hamiltonian is not Hermitian in this sector, and that the continuation not of $f_{\rm BW}(E)$ but of $f_{ PT}(E)$ between real $E$ and the complex poles is both needed and valid if the poles are to be associated with eigenvalues of the scattering process.  

\section{Final Comments}
\label{S6}

\begin{figure}[H]
\centering
\includegraphics[scale=0.3]{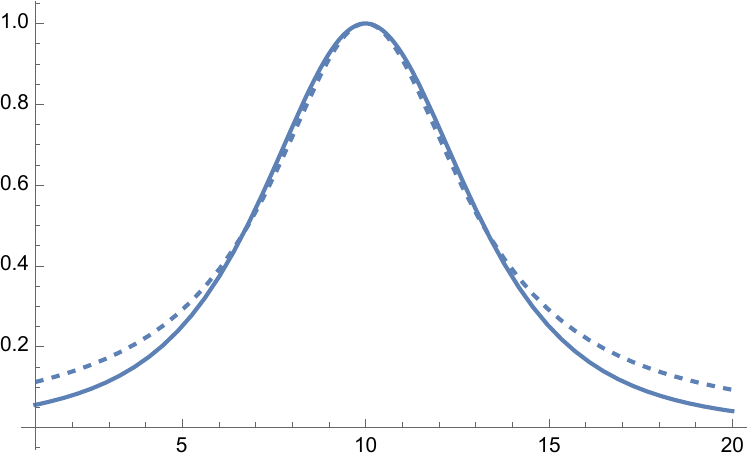}~~~\includegraphics[scale=0.3]{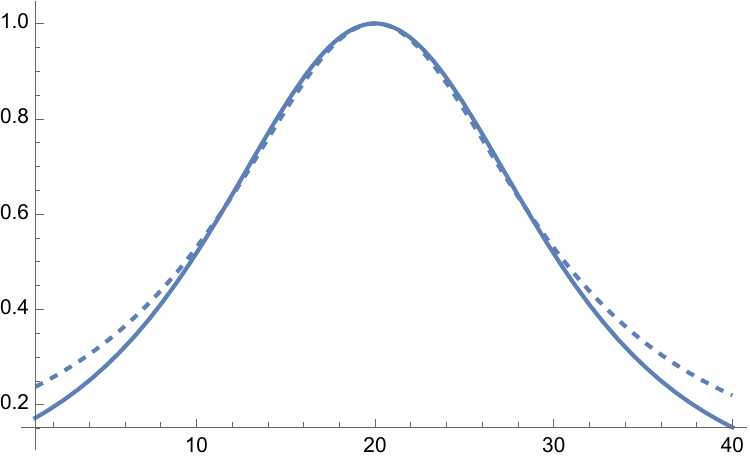}~~~
\includegraphics[scale=0.3]{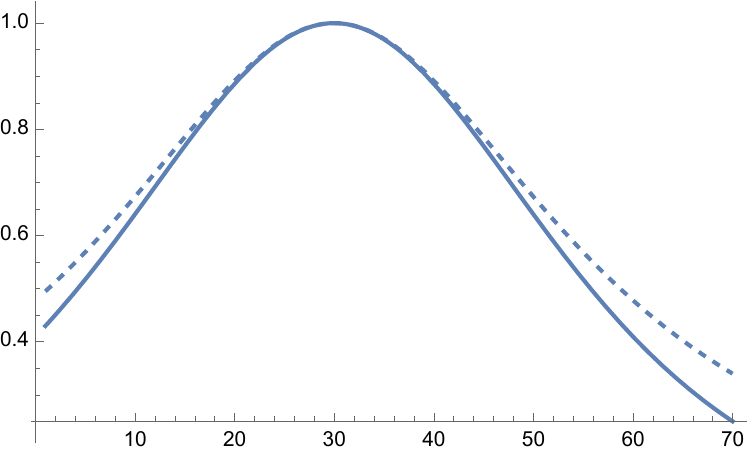}
\caption{$\vert f_{PT}(E)\vert^2$ (solid curve)  and $\vert f_{\rm BW}(E)\vert^2$ (dashed curve) for $E_2= E_1=10, 20,30$, 
$\Gamma_2=5,16,40$, $\Gamma_1=3.2,10.6,28.1.$}
\label{fit}
\end{figure}

In regard to the use of   $\vert f_{PT}(E)\vert^2$ for experimental data in a closed system (such as $\pi^++p$ elastic scattering), in Fig. \ref{fit}  we plot its contribution to a cross-section as a function of energy in three typical cases: $E_2=10$, $\Gamma_2=5$, $E_2=20$, $\Gamma_2=16$, $E_2=30$, $\Gamma_2=40$. We have normalized $\vert f_{PT}(E)\vert^2$ to $\vert f_{PT}(E_2)\vert^2=1$, as could be achieved by replacing  $V=i\sigma_2$ by $V=i\sigma_2/2$ in (\ref{4.8}) since this would leave $M_2^{\dagger}=VM_2V^{-1}$ unchanged. To compare with  $\vert f_{\rm BW}(E)\vert^2$ we also plot $\vert f_{\rm BW}(E)\vert^2$  for $E_1=E_2$ and vary $\Gamma_1$, and  get agreement with  $\vert f_{PT}(E)\vert^2$ when $\Gamma_1$ respectively equals $3.2$,  $10.6$ and $28.1$ in the three cases. In the square well case there would also be a background contribution coming from the continuum modes, though as we noted in Sec. \ref{S3}, the $PT$ resonances are distributed quite far from the continuum resonances, with the $PT$ resonances thus occurring where the continuum background is not substantial. As we see from the figures $\vert f_{PT}(E)\vert^2$ and $\vert f_{\rm BW}(E)\vert^2$ agree remarkably well. Thus we should now recognize that what were thought to be Breit-Wigner fits are instead $PT$ fits. This would not affect an extracted value for the position of a resonance peak at $E_2=E_1$, but would require a reevaluation of a fitted resonance width since $\Gamma_2$ is not equal to the fitted  $\Gamma_1$. 

To conclude we see that we have established  the centrality of antilinear symmetry and pseudo-Hermiticity to the square well scattering sector, and in the square well see confirmation of our general contention \cite{Mannheim2018a} that antilinearity is a more general guideline for quantum theory than Hermiticity.
Finally,  we note that any system for which the finite-depth square well is a prototype (going back to as early as the first discussions of the finite-depth square well itself in quantum mechanics but excluding systems with an infinitely high potential barrier) is actually an antilinear theory rather than a Hermitian theory, with $PT$-type theories thus having existed long before they were identified as such.

\noindent
Acknowledgment: 
The author wishes to acknowledge helpful comments from a referee.
\smallskip

\noindent
Data Availability Statement:  Data sharing is not applicable to this article as the only data used are presented in this paper as  Mathematica plots.

\medskip
\appendix
\numberwithin{equation}{section}
\setcounter{equation}{0}
\section{Constraints on $k_I$}

\subsection{Outgoing waves}

From the outgoing wave tangent equation given in (\ref{3.10c}) we have
\begin{align}
ka\sin(Ka)+iKa\cos(Ka)=0. 
\label{A.1}
\end{align}
On setting $a=1$, we have
\begin{align}
(k_R+ik_I)[\sin(K_R)\cosh(K_I)+i\cos(K_R)\sinh(K_I)]+i(K_R+iK_I))[\cos(K_R)\cosh(K_I)-i\sin(K_R)\sinh(K_I)]=0.
\label{A.2}
\end{align}
Thus we obtain
\begin{align}
&k_R\sin(K_R)\cosh(K_I)-k_I\cos(K_R)\sinh(K_I)+K_R\sin(K_R)\sinh(K_I)-K_I\cos(K_R)\cosh(K_I)=0,
\nonumber\\
&k_I\sin(K_R)\cosh(K_I)+k_R\cos(K_R)\sinh(K_I)+K_I\sin(K_R)\sinh(K_I)+K_R\cos(K_R)\cosh(K_I)=0,
\label{A.3}
\end{align}
so that
\begin{align}
\frac{\sin(K_R)}{\cos(K_R)}=\frac{k_I\sinh(K_I)+K_I\cosh(K_I)}{K_R\sinh(K_I)+k_R\cosh(K_I)}=-\frac{(k_R\sinh(K_I)+K_R\cosh(K_I))}{(K_I\sinh(K_I)+k_I\cosh(K_I))},
\label{A.4}
\end{align}
\begin{align}
(k_IK_I+k_RK_R)(\cosh^2(K_I)+\sinh^2(K_I))+(k_I^2+K_I^2+k_R^2+K_R^2)\sinh(K_I)\cosh(K_I)=0,
\label{A.5}
\end{align}
\begin{align}
\tanh(2K_I)=-\frac{2(k_IK_I+k_RK_R)}{(k_I^2+K_I^2+k_R^2+K_R^2)}.
\label{A.6}
\end{align}

Now from (\ref{3.16e}) it follows that $k_Rk_I=K_RK_I$, and none of these terms can be zero as long as $\Gamma_2$ is not zero. Thus from (\ref{A.6}) we obtain
\begin{align}
\frac{\tanh(2K_I)}{2K_I}=-\frac{(k_I+k_RK_R/K_I)}{(k_I^2+K_I^2+k_R^2+K_R^2)}=-\frac{(k_I^2+K_R^2)}{k_I(k_I^2+K_I^2+k_R^2+K_R^2)}.
\label{A.7}
\end{align}
Since the left-hand side of (\ref{A.7}) is an even function of $K_I$ and is positive for any sign of $K_I$, it follows that $k_I$ is uniquely negative no matter what the signs of $k_R$, $K_R$ or $K_I$. Thus the only solutions to (\ref{A.1}) are those in which $k_I$ is negative. Then with (\ref{3.16e}) also giving $\hbar^2k_Rk_I/m=-\Gamma_2$, for any outgoing $e^{ik_Rr-k_Ir}e^{-iE_2t-\Gamma_2t}$ mode with $k_R$ positive, exponential decay in time must be accompanied by exponential growth in $r$, viz. expressly a Gamow vector. 

Bound states have $k_I>0$ and real energy $E<V_0$, and therefore do not fall into the outgoing $\Gamma_2\neq 0$ class with $k_I<0$. They satisfy (\ref{A.6}) identically by having $K_I=0$, $k_R=0$, so that  $k_Rk_I=K_RK_I=0$. 

From (\ref{A.3}) we also obtain 
\begin{align}
\frac{\sinh(K_I)}{\cosh(K_I)}=-\frac{(k_R\sin(K_R)-K_I\cos(K_R))}{(K_R\sin(K_R)-k_I\cos(K_R))}=-\frac{(k_I\sin(K_R)+K_R\cos(K_R))}{(K_I\sin(K_R)+k_R\cos(K_R))},
\label{A.8}
\end{align}
\begin{align}
(k_RK_I-k_IK_R)(\sin^2(K_R)-\cos^2(K_R))+(k_I^2-K_I^2-K_R^2+k_R^2)\sin(K_R)\cos(K_R)=0,
\label{A.9}
\end{align}
\begin{align}
\tan(2K_R)=\frac{2(k_RK_I-k_IK_R)}{(k_I^2-K_I^2-K_R^2+k_R^2)}=\frac{2K_R(k_Rk_IK_I/K_R-k_I^2)}{k_I(k_I^2-K_I^2-K_R^2+k_R^2)}=\frac{2K_R(K_I^2-k_I^2)}{k_I(k_I^2-K_I^2-K_R^2+k_R^2)}.
\label{A.10}
\end{align}
Thus we obtain 
\begin{align}
\frac{\tan(2K_R)}{2K_R}=\frac{(K_I^2-k_I^2)}{k_I(k_I^2-K_I^2-K_R^2+k_R^2)}.
\label{A.11}
\end{align}
While $\tan(2K_R)/2K_R)$ is an even function of $K_R$, it is not positive definite. Even with positive $K_R$ it could still be negative, just as the right-hand side of (\ref{A.11}) can be. It thus leads to no constraint of the type obtained from (\ref{A.7}) that requires $k_I$ to be negative. 

For bound states  the first equation in (\ref{A.10}) reduces to $\tan(2K_R)=-2k_IK_R/(k_I^2-K_R^2)$, from which it follows that $\tan(K_R)=-K_R/k_I$, with $\beta=ka\tan(Ka)/Ka=-i$ then reducing to $\tan(K_R)=-K_R/k_I$ just as required. However, for bound states this does not fix the sign of $k_I$, and while it is positive in outgoing modes for bound states (exponential radial falloff), it can also be negative, to give anti-bound states (exponential radial growth). Both bound and anti-bound states are $\tan\delta =-i$ poles in the scattering amplitude. While not in and of themselves square integrable, anti-bound states can play a role in scattering experiments, especially at low energy.

\subsection{Ingoing waves}

From (\ref{3.8c}) ingoing waves with $e^{-ikr}$ have $\tan\delta-i=0$, to thus have $\beta=+i$ and yield 
\begin{align}
ka\sin(Ka)-iKa\cos(Ka)=0.
\label{A.12}
\end{align}
On setting $a=1$, we have
\begin{align}
(k_R+ik_I)[\sin(K_R)\cosh(K_I)+i\cos(K_R)\sinh(K_I)]-i(K_R+iK_I))[\cos(K_R)\cosh(K_I)-i\sin(K_R)\sinh(K_I)]=0.
\label{A.13}
\end{align}
Thus we obtain
\begin{align}
&k_R\sin(K_R)\cosh(K_I)-k_I\cos(K_R)\sinh(K_I)-K_R\sin(K_R)\sinh(K_I)+K_I\cos(K_R)\cosh(K_I)=0,
\nonumber\\
&k_I\sin(K_R)\cosh(K_I)+k_R\cos(K_R)\sinh(K_I)-K_I\sin(K_R)\sinh(K_I)-K_R\cos(K_R)\cosh(K_I)=0,
\label{A.14}
\end{align}
so that
\begin{align}
\frac{\sin(K_R)}{\cos(K_R)}=\frac{(k_I\sinh(K_I)-K_I\cosh(K_I))}{(-K_R\sinh(K_I)+k_R\cosh(K_I))}=\frac{(-k_R\sinh(K_I)+K_R\cosh(K_I))}{(-K_I\sinh(K_I)+k_I\cosh(K_I))},
\label{A.15}
\end{align}
\begin{align}
-(k_IK_I+k_RK_R)(\cosh^2(K_I)+\sinh^2(K_I))+(k_I^2+K_I^2+k_R^2+K_R^2)\sinh(K_I)\cosh(K_I)=0,
\label{A.16}
\end{align}
\begin{align}
\tanh(2K_I)=\frac{2(k_IK_I+k_RK_R)}{(k_I^2+K_I^2+k_R^2+K_R^2)}.
\label{A.17}
\end{align}

Whether $\Gamma_2$ is positive or  negative  from (\ref{3.16e}) it follows that $k_Rk_I=K_RK_I$, and none of these terms can be zero as long as $\Gamma_2$ is not zero. Thus from (\ref{A.17}) we obtain
\begin{align}
\frac{\tanh(2K_I)}{2K_I}=\frac{(k_I^2+K_R^2)}{k_I(k_I^2+K_I^2+k_R^2+K_R^2)}.
\label{A.18}
\end{align}
Since the left-hand side of (\ref{A.18}) is an even function of $K_I$ and is positive for any sign of $K_I$, it follows that $k_I$ is uniquely positive no matter what the signs of $k_R$, $K_R$ or $K_I$. Thus the only solutions to (\ref{A.18}) are those in which $k_I$ is positive. Then with (\ref{3.16e}) giving $\hbar^2k_Rk_I/m=-\Gamma_2$, for any ingoing $e^{-ik_Rr+k_Ir}e^{-iE_2t+\Gamma_2t}$ mode with $k_R$ positive and $\Gamma_2$ negative, ingoing exponential growth in time must be accompanied by exponential growth in $r$.
Thus both outgoing and ingoing mode solutions to the square well problem have exponential growth in space. (The definition of $k_I$ discussed here for ingoing modes is opposite of that used earlier for an ingoing time-reversed conjugate $\psi^{(c)}_{PT}$, but for either case the effect is still one of exponential growth.)

In the ingoing sector bound states are replaced by anti-bound states that have $k_I>0$ and real energy $E<V_0$, and do not fall into the ingoing $\Gamma_2\neq 0$ class with $k_I>0$. They satisfy (\ref{A.17}) identically by having $K_I=0$, $k_R=0$, so that  $k_Rk_I=K_RK_I=0$. 

From (\ref{A.14}) we also obtain 
\begin{align}
\frac{\sinh(K_I)}{\cosh(K_I)}=\frac{(k_R\sin(K_R)+K_I\cos(K_R))}{(K_R\sin(K_R)+k_I\cos(K_R))}=\frac{(k_I\sin(K_R)-K_R\cos(K_R))}{(K_I\sin(K_R)-k_R\cos(K_R))},
\label{A.19}
\end{align}
\begin{align}
(k_RK_I-k_IK_R)(\sin^2(K_R)-\cos^2(K_R))-(k_I^2-K_I^2-K_R^2+k_R^2)\sin(K_R)\cos(K_R)=0,
\label{A.20}
\end{align}
\begin{align}
\tan(2K_R)=-\frac{2(k_RK_I-k_IK_R)}{(k_I^2-K_I^2-K_R^2+k_R^2)}=-\frac{2K_R(k_Rk_IK_I/K_R-k_I^2)}{k_I(k_I^2-K_I^2-K_R^2+k_R^2)}=-\frac{2K_R(K_I^2-k_I^2)}{k_I(k_I^2-K_I^2-K_R^2+k_R^2)}.
\label{A.21}
\end{align}
Thus we obtain 
\begin{align}
\frac{\tan(2K_R)}{2K_R}=-\frac{(K_I^2-k_I^2)}{k_I(k_I^2-K_I^2-K_R^2+k_R^2)}.
\label{A.22}
\end{align}
While $\tan(2K_R)/2K_R)$ is an even function of $K_R$, it is not positive definite. Even with positive $K_R$ it could still be negative, just as the right-hand side of (\ref{A.22}) can be. It thus leads to no constraint of the type obtained from (\ref{A.18}) that requires $k_I$ to be positive. 

For bound states the first equation in (\ref{A.21}) reduces to $\tan(2K_R)=2k_IK_R/(k_I^2-K_R^2)$, from which it follows that $\tan(K_R)=K_R/k_I$, with $\beta=ka\tan(Ka)/Ka=i$ then reducing to $\tan(K_R)=K_R/k_I$ just as required. However, for bound states this does not fix the sign of $k_I$, and while it is negative in ingoing modes for bound states (exponential radial falloff), it can also be positive, to give anti-bound states (exponential radial growth). Both bound and anti-bound states are $\tan\delta =+i$ poles in the time reversed scattering amplitude.

\end{document}